\documentclass[useAMS,usenatbib]{mnras}
\usepackage[utf8]{inputenc}
\usepackage{amssymb}
\usepackage{amsmath}
\usepackage{graphicx}
\usepackage{multirow}
\usepackage{tabularx}
\usepackage{siunitx}

\def\Chandra{\mbox{\textit{Chandra}}}
\def\Swift{\mbox{\textit{Swift}}}
\def\HST{\mbox{\textit{HST}}}

\def\obj{\mbox{NGC\,5474~X-1}}

\begin{document}
\title[NGC 5474 X-1: a neutron star ULX]
{NGC\,5474 X-1: a neutron star ULX in an old stellar cluster?}

\author[K. Atapin et al.]
{Kirill Atapin,$^{1,2}$
Alexander Vinokurov,$^{2}$\thanks{E-mail: vinokurov@sao.ru} 
Arkadiy Sarkisyan,$^{2}$ 
Yulia Solovyeva,$^{2}$
\newauthor
Alexander Kostenkov,$^{2}$
Aleksei Medvedev,$^{2}$
Sergei Fabrika$^{2}$\\
$^1$ Sternberg Astronomical Institute, Moscow State University, Universitetsky pr., 13, Moscow, 119991, Russia\\
$^2$ Special Astrophysical Observatory, Nizhnij Arkhyz, 369167, Russia}
\pagerange{\pageref{firstpage}--\pageref{lastpage}}
\pubyear{2023}
\date{Accepted 2023 November 28. Received 2023 November 23; in original form 2023 September 29}

\label{firstpage} 
\maketitle

\begin{abstract}
We present the optical and X-ray study of the ultraluminous X-ray source (ULX) NGC\,5474~X-1. The X-ray spectrum taken during the bright state of the source ($L_X \sim 2\times10^{40}$ egs/s) shows signatures of a broad absorption line at $\simeq 8$~keV which may be a cyclotron resonant scattering feature. This implies that this system may host a neutron star with a magnetic field $\sim 10^{12}$~G. The first observation of this area with the Hubble Space Telescope (HST) carried out 14 months later revealed that the source was bright in the optical range as well. The subsequent observations have shown that the source faded in both ranges (more than $2.8^m$ in the U band and by a factor of 50-100 in X-rays) and has never become bright again. Deeper HST observations made it possible to impose constraints on the donor star spectral class and mass ($<\,7$~M$_\odot$), as well as to identify a stellar cluster of about 1 Gyr, the centre of which is located at a projected distance of $\simeq 2$~pc from NGC\,5474~X-1. The ULX can be a member of this old cluster, however, the presence of stars with ages of $\sim10$~Myr within 300~pc around the ULX does not completely exclude the possibility that it is just an accidental projection.

\end{abstract}


\section{Introduction}

Ultraluminous X-ray sources are defined as extragalactic, off-nuclear, point-like objects with X-ray luminosity exceeding $\sim2\times10^{39}$~erg/s, the Eddington luminosity for a typical black hole in the Milky Way \citep{Fabrika2021review,King2023review}. 
Modern catalogues of ULX candidates include  about a thousand sources \citep{Kovlakas2020ULXcat,Walton2022ULXcat, Tranin2023} 
and this number is still growing. Since the discovery, several scenarios were proposed to explain the nature of ULXs. Firstly, it was suggested that such objects could be accreting intermediate-mass black holes with masses $10^2$--$10^5$\,M$_\odot$ \citep{Colbert1999ULXIMBH}. The alternative scenario proposed supercritical accretion onto stellar mass black holes \citep{FabrikaMescheryakov2001,King2001ULX,Poutanen2007} or magnetized neutron stars \citep{Mushtukov2015ULXP,Kawashima2016ULXpulsmodel,KingLasota2016,Chashkina2019} to explain such high X-ray luminosities of the ULXs.  Also, it was shown that young rotation-powered pulsars may theoretically reach sufficiently high luminosities \citep{Perna2004CrabULX,MedvedevPoutanen2013}.

It is now clear that at least part of ULXs are accreting neutron stars (NS-ULXs). Most of them were revealed by the detection of coherent pulsations of their X-ray fluxes (ultraluminous X-ray pulsars, hereafter ULXP, \citealt{Bachetti2014Nat,Furst2016pulsP13,Israel2017SciNGC5907,Israel2017pulsP13,Carpano2018NGC300,Sathyaprakash2019pulsNGC1313X2,RodriguezCastillo2020pulsM51ulx7,Townsend2017SMCX3,Chandra2020ULXPinSMC,Vasilopoulos2020ULXPinSMC,Quintin2021NGC7793ULX4}). However in the case of M51~ULX-8 the presence of a neutron star follows from the discovery of a cyclotron resonant scattering feature (CRSF) in its X-ray spectrum \citep{Brightman2018M51ULX8NS}, the absorption line that appears due to transitions of charged particles between Landau levels. The observed line energy $\approx4.5$~keV corresponds to the magnetic field strength of either $10^{12}$~G or $10^{15}$~G depending on electron or proton nature of the line \citep{Brightman2018M51ULX8NS,Middleton2019M51ULX8}. Despite such a strong magnetic field and high accretion rate, this source does not demonstrate pulsations. Nevertheless, M51~ULX-8 may still be analogue to ULXPs, with pulsations invisible for the Earth observer due to geometric factors~\citep{KingLasota2020}. Also, a CRSF at $\sim 13$~keV has been suspected in the confirmed ultraluminous pulsar NGC\,300~ULX1~\citep{Walton2018NGC300line} and at $\approx8.5$~keV in the hyperluminous source in the NGC\,4045 galaxy \citep{Brightman2022NGC4045CRSF}.

Besides the pulsations, ULXPs demonstrate hard spectra (with spectral index $\Gamma\sim 1$, \citealt{Pintore2017,Walton2018puls}) and very strong variability. For example, M82~X-2, NGC\,7793 P13, and NGC\,5907 X-1 show `off' states in which their fluxes drop by factors of $\sim100$ \citep{Motch2014NaturP13,Walton2015NGC5907,Brightman2016M82X2}. It was proposed that such off-states may be caused by transitions into the propeller regime in which the magnetic field pressure begins to suppress accretion \citep{Tsygankov2016propellerM82}.

In the optical range, ultraluminous X-ray sources have a wide variety of observational properties. The absolute magnitudes of the ULX optical counterparts lie in the range $M_V = -3\div -8$ \citep{Tao2011counterparts,Vinokurov2018,Vinokurov2020a}. The most bright counterparts mainly have blue, power-law spectral energy distributions (SEDs), whereas the less bright sources show SEDs consistent with spectra of A-G class stars \citep{Avdan2016,Avdan2019}. This may suggest that the spectra of the brightest counterparts are fully dominated by optical emission of the supercritical accretion disc but as the disc contribution decreases, the donor emission becomes more prominent \citep{Vinokurov2018}. Most of the ULXs that are not strongly variable in X-rays show a modest variability in the optical range as well (within 0.2 mag in the V band, \citealt{Tao2011counterparts}). In contrast to them, the transient ULXs along with the confirmed ULXPs show the optical variability $\gtrsim 1$~mag in the visual bands which has been found to correlate with the X-ray flux in some cases \citep{Soria2012M83tULX,Motch2014NaturP13,Villar2016Sn2010da,Vinokurov2020b}.

The optical spectroscopy also gave many important results. It has been shown that there are powerful outflows in the ULX binary systems in a form of relatively slow winds (500--1600~km/s) which more likely come from the surface of the supercritical accretion disc \citep{Fabrika2015}. However, this outflows seem to differ from the ultrafast outflows discovered in X-rays \citep{Pinto2016Nat,Pinto2017NGC55ULX,Kosec2018NGC300UFO} in both the velocities and the outflow rates \citep{Kostenkov2020}. Another important result of the optical spectroscopy is the detection of spectral lines of donor stars. The first identified donor was the Wolf-Rayet star in M101~ULX-1 discovered by \cite{Liu2013NaturM101ULX1}. Then the signatures of a B9\,Ia star (the Balmer jump together with the absorption lines of hydrogen, helium and some other elements) were found in the spectra of the ultraluminous pulsar NGC\,7793~P13 \citep{Motch2014NaturP13}. Currently, the most common type of ULX donors is a red supergiant (6 systems, \citealt{Heida2015NGC253,Heida2016fiveULXs,Heida2019NGC300,Lopez2020}).

Much of useful information can be obtained from the study of the ULX environment. Many of optically studied sources are surrounded by shock-ionized bubbles with sizes from a few to hundreds parsecs \citep{PakullMirioni2003,Lehmann2005HoIIX1,Griese2012NGC5408X1}. These bubbles are thought to be excited by the interaction of the ULX winds (and also may be in some cases jets) with surrounding gas, as well as by ionisation with X-ray and UV emission \citep{Abolmasov2007}. Many ULXs residing in star-forming galaxies are associated with compact stellar clusters or super-star clusters and should have a physical relation to them. Such a relation has been firmly proven by \cite{Poutanen2013AntennaeULX} on the example of the ULX population of the galaxies NGC4038/NGC4039. Nevertheless, even in galaxies with high star-formation rate, there are ULXs that are located far from young stars, as, for example, the ULX in M83~\citep{Soria2012M83tULX}. Moreover, it has been shown that in elliptical galaxies ULXs may even be inside globular clusters \citep{Irwin2004ULXsEarlyTypeGal}. However, the luminosities of such old objects usually do not exceed 2--3$\times 10^{39}$~erg/s \citep{Sarazin2000XrayFaintGal,KimFabbiano2010}. 

In this work, we report possible detection of a broad absorption line, which may be a CRSF, in the X-ray spectrum of \obj. This source located in a dwarf spiral galaxy at 6.8~Mpc \citep{DrozdovskyKarachentsev2000} was marked as ULX by \cite{Swartz2011ULXcat} and studied by \cite{Avdan2016} who identified its possible optical counterpart. Our more careful analysis of the archival data revealed an optical outburst of the ULX which allowed us to refine the source position and to make the conclusion that the counterpart by \cite{Avdan2016} is actually the host (or background) stellar cluster of this ULX. Using additional \textit{Hubble Space Telescope} (\HST) observations, we determined the age of the cluster and provided new constraints to the spectral class and the mass of the ULX donor star. The paper is organised as follows. In section~2 we describe available X-ray observations and present the spectra and the long-term light curve. Since the most accumulated spectrum is affected by pile-up, we involved Bayesian methods to choose between different spectral models. The section 3 is devoted to the optical data. Here we carried out the photometry of the ULX counterpart, the stellar cluster and their surrounding stars and then fitted the obtained spectral energy distributions (SEDs) with different models. In section~4 we discuss our results.

\section{X-ray data}
\label{sec:xrays}
In X-rays, the source was observed twice with the {\it Chandra X-ray observatory} and 20 times with the {\it Neil Gehrels Swift observatory}. The \Chandra\ observations were carried out on 2006 September 10 (ObsID\,7086, MJD = 53988) and 2007 December 3 (ObsID\,9546, MJD = 54437); their exposure times are 1.7\,ks and 30\,ks, respectively. Both observations are taken in the `faint' mode with the target placed at the nominal aim-point on the S3 chip. The \Swift\ observations of the NGC\,5474 galaxy were taken between April 2011 and January 2023 but most of them fall on the autumn of 2012, a total exposure time is $\simeq 25$\,ks. Our preliminary investigation of the data has shown that the source was at ULX luminosities only in the two \Chandra\ observations. By the \Swift\ epoch, it became much fainter and has never returned to the bright state (see Sec.\,\ref{sec:xraylc}).

The reduction of the Chandra data was done using the \textsc{ciao}\,4.13 package with CALDB\,4.9.6. We created new level-2 event files with the \texttt{repro} script following the standard procedure recommended by the \Chandra\ X-ray Center. The \Swift/XRT event files were reprocessed with the task \texttt{xrtpipeline} (version 0.13.5) of \textsc{heasoft}\,6.28. A spectral analysis was performed with \textsc{xspec}\,12.11.1 \citep{Arnaud1996xspec}.

\begin{figure*}
\includegraphics[angle=270,width=0.45\textwidth]{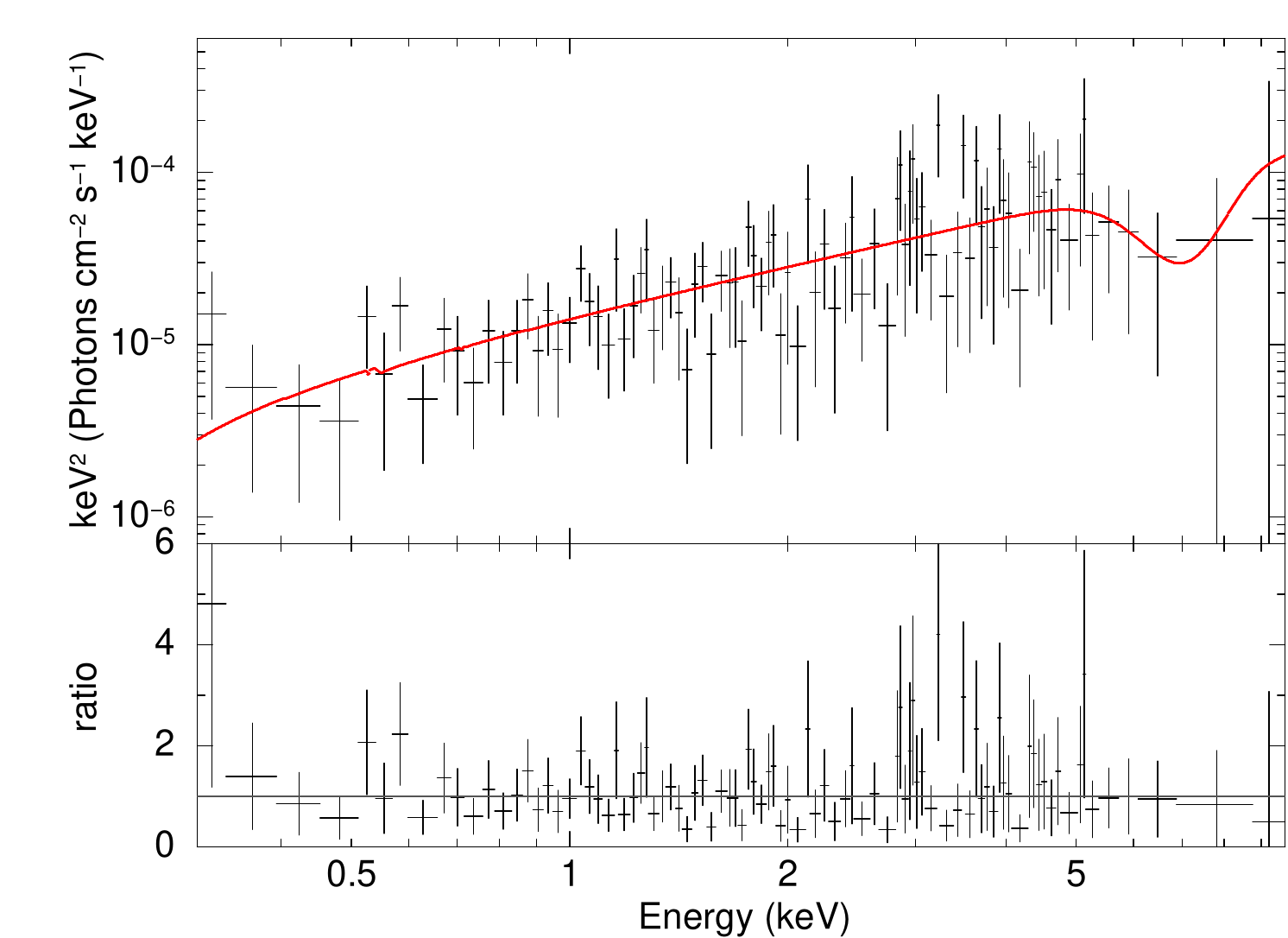}
\includegraphics[angle=270,width=0.45\textwidth]{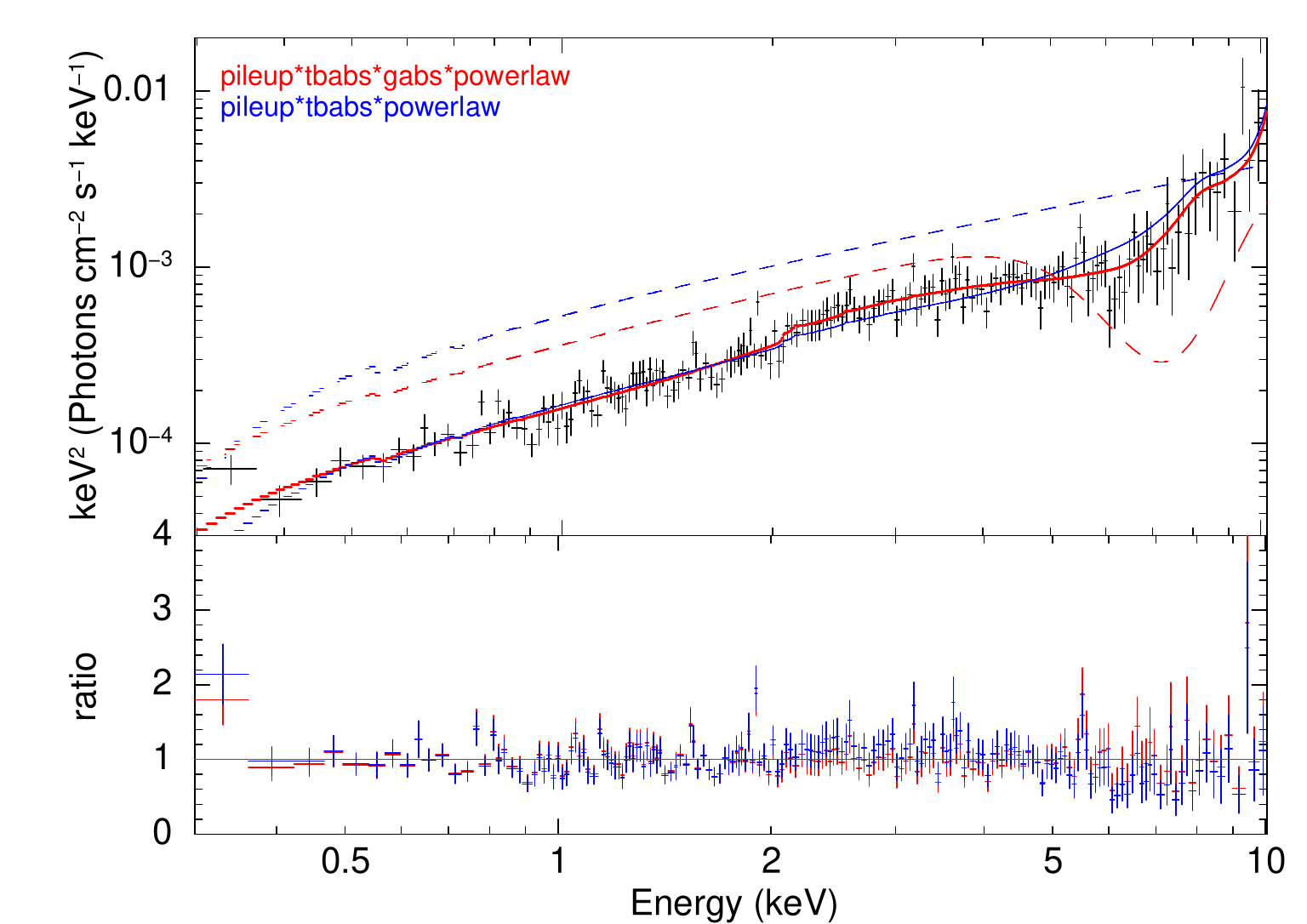}
\caption{Unfolded spectra from the \Chandra/ACIS observation ObsID\,9546 suffering from moderate pile-up. \textit{Left:} the spectrum extracted from the annular aperture that excludes most of piled events, fitted by an absorbed power law with a Gaussian absorption line model. \textit{Right:} the spectrum from a circular aperture fitted by an absorbed power law with (red) and without (blue) the Gaussian absorption component, convolved with the \texttt{pileup} model (solid lines). The same models with pile-up components disabled for illustrative purpose are shown by the dashed lines. Pile-up redistributes soft photons to higher energies which produces a sharp rise above 6~keV in the observed spectrum and hides the absorption line. Although the line is not visible directly, its presence is essential to fit the data (see text). The spectra were fitted with the Cash statistic and rebinned only for visual clarity.}
\label{fig:spectra_xspec}
\end{figure*}

\subsection{Spectral analysis}
\label{sec:xray_spec}

We found that both Chandra observations of \obj\ are affected by the pile-up effect. Pile-up occurs when two or more photons fall in a single detector cell between two consecutive frame readouts (see ``The Chandra ABC Guide to Pileup''\footnote{https://cxc.harvard.edu/ciao/download/doc/pileup\_abc.pdf}, hereafter the pile-up guide). The detector cannot resolve such photons and records them as a single event of higher energy. This distorts an observed spectrum making it harder and also leads to underestimation of a source flux. This occurs, among other things, because some part of the piled events generates bad grades even if they are produced by valid X-ray photons (so-called `grade migration'). Such bad graded events are indistinguishable from true bad events produced by charged particles, so they are doomed to be filtered out at stages of the primary data reduction. The pile-up fraction depends on the count rate and may be high if the source is too bright for the chosen CCD readout speed.

To estimate the pile-up fraction in our observations, we used the \textsc{ciao} tool \texttt{pileup\_map} which converts the image to the units of the average number of counts per ACIS frame. We found that the brightest pixel of X-1 in ObsID\,9546 has 0.46~counts/frame which corresponds to the pile-up fraction of $\sim 20$\% according to the blue curve in Fig.~3 of the pile-up guide (for the grade migration parameter $\alpha=0.5$). The pile-up fraction of ObsID\,7086 having 0.3~counts/frame in the brightest pixel is estimated as $\sim10$\%. In both cases the pile-up may be considered as moderate.

There are two common approaches to deal with observations suffering from pile-up. First of all, one can exclude the core region of the PSF by using an annular aperture. This rejects most of piled events preventing the distortion of the spectrum but also significantly reduces the number of good events available for analysis. Alternatively, one can extract the spectrum from a full (circular) aperture and modify the fitting model by a special convolution model component to account for spectral distortions produced by pile-up. Such a convolution model for the Chandra CCD detectors is developed by \cite{Davis2001pileup}. 
Below we tried both approaches.

For the longer observation ObsID\,9546, using the \texttt{specextract} task, we extracted a spectrum from a circular aperture of $3''$ radius and a number of spectra from annular apertures with an outer radius of $3''$ and an inner radius varied from $1.0''$ to $2.0''$. All the apertures were centred at the (same) position of the PSF centroid calculated with the \texttt{dmstat} tool in \textsc{ds9}. The background was extracted from a big annulus around X-1 free of other sources. Then all the spectra were grouped to have a minimum of 1 count per bin to use the C-statistic \citep{Cash1979}. Events outside the energy range 0.3--10~keV were ignored.

Fig.~\ref{fig:spectra_xspec}a shows the spectrum from the annular aperture with the inner radius $1.3''$ which gave the maximal signal-to-noise ratio with almost absent piled events. The spectrum demonstrates a deepening near 7~keV. We fitted this spectrum with the \textsc{xspec} model \texttt{tbabs*gabs*powerlaw} consisting of a Gaussian absorption line and a power-law continuum modified for interstellar absorption. The hydrogen column density of the absorption component was restricted to vary to be above the Galactic value $1.9\times10^{20}$~cm$^{-2}$ estimated via the web tool \texttt{Nh}\footnote{https://heasarc.gsfc.nasa.gov/cgi-bin/Tools/w3nh/w3nh.pl} 
based on HI4PI map \citep{BenBekhti2016HI4PImap}. The best-fitting model parameter are: the hydrogen column density $N_H \approx 2\times 10^{20}$~cm$^{-2}$, the power-law spectral index $\Gamma=1.05\pm0.11$, the line centroid energy $L_E \approx 7$~keV, the width $L_\sigma \approx 0.9$~keV and the line strength $L_{st}\approx 2.8$; the fit statistic is $C=127.3$ for 176 degrees of freedom (d.o.f). Due to low counting statistics of this spectrum it was possible to determine only the lower boundaries of the 1-$\sigma$ confidence intervals for the line parameters: $L_E >6$~keV, $L_\sigma > 0.4$~keV and $L_{st} > 0.8$.

\begin{table*}
\begin{minipage}{170mm}
\caption{Results of the Bayesian analysis of the Chandra ObsID\,9546 spectrum of \obj.}
\label{tab:spec_results}
\begin{tabular}{lcccccccccc} \hline\hline
Model & $\alpha^a$ & $N_H$, & $L_E$, & $L_{\sigma}$, & 
$L_{\rm st}^b$ & $T_{\rm in}$, & $\Gamma$ & $E_{\rm cut}$, & $\Delta \log\,Z^c$ \\
\texttt{pileup*tbabs*} & & $\times10^{20}$\,cm$^2$ & keV & keV & & keV & & keV & \\
 \hline
Line*(Disc+PL) & $0.54\pm0.17$ & $3.2\pm 1.2$ & 
$7.9\pm1.6$ & $1.8\pm0.7$ & $12\pm5$ & 
$3.9\pm2.4$ & $1.8\pm0.7$ & --- & 0.0 \\
Line*PL & $0.55\pm0.16$ & $3.1\pm 1.1$ & 
$8.0\pm0.9$ & $1.7\pm0.5$ & $13\pm4$ & 
--- & $1.17\pm0.11$ & -- & $-0.01$ \\
Line*CutPL & $0.60\pm0.17$ & $2.9\pm0.9$ &
$7.9\pm0.9$ & $1.5\pm0.6$ & $13\pm5$ &
--- & $1.08\pm0.11$ & $17\pm7$ & $-0.04$ \\ \hline

(Disc+PL) & $0.57\pm0.18$ & $3.2\pm1.2$ & 
--- & --- & --- &
$1.57\pm0.27$ & $2.6\pm0.3$ & --- & $-1.43$ \\
Line*Disc & $0.60\pm0.19$ & $2.5\pm0.6$ & 
$2.2_{-1.4}^{+2.4}$ & $2.1\pm0.8$ & $9\pm 6$ & 
$1.1\pm0.5$ & --- & --- & $-2.18$ \\
Disk & $0.61\pm0.17$ & $2.2\pm0.3$ &
 --- & --- & --- & 
$1.27\pm0.24$ & --- & --- & $-3.05$ \\
CutPL & $0.72\pm0.16$ & $2.7\pm0.8$ & 
--- & --- & --- & 
--- & $0.75\pm0.13$ & $2.5\pm0.6$ & $-3.16$ \\
PL & $0.40\pm0.26$ & $2.7\pm0.9$ & 
--- & --- & --- & 
--- & $0.98\pm0.08$ & --- & $-9.53$ \\ \hline
\end{tabular}
{\it Notes.} `Disc', `PL', `CutPL' and `Line' refer to the  
\texttt{diskbb}, \texttt{powerlaw}, \texttt{cutoffpl} and \texttt{gabs} xspec models. $^a$Grade migration parameter of the \texttt{pileup} model. $^b$Strength of the absorption line. $^c$Log-evidence of the model relative the best one. The models below the horizontal line should be ruled out. The shown parameter values represent the mean and 1-$\sigma$ credible interval of the corresponding posterior distribution.
\end{minipage}
\end{table*}

The spectrum from the circular aperture was fitted by the same model with the \texttt{pileup} component appended. The \texttt{pileup} model \citep{Davis2001pileup} has a number of parameters two of which the user can leave free to vary: the grade migration parameter $\alpha$ and the PSF fraction $F$. The first one represents the probability for piled events to produce good grades and ranges from 0 to 1. The pile-up guide states that the preferred value of $\alpha$ is unknown, however, values near the extremes should be considered as unrealistic, and there are some observational hints on $\alpha \gtrsim 0.5$. The second one is the fraction of accumulated events which will be treated by the pile-up model. Assuming that pile-up affects mainly the central $3\times 3$~pix region of a PSF, \cite{Davis2001pileup} estimated $F$ as about 95\% for a point on-axis source. In our case, using the \texttt{ecf\_calc} tool, we found that the circumcircle of the $3\times 3$~pix region gather $\approx93$\% of total counts accumulated by the $3''$ aperture. Therefore, we restricted $F$ to be 0.9 or above. The parameters of the Gaussian absorption, power law and $\alpha$ were allowed to vary within their default ranges. Fitting the model to the spectrum from the circular aperture, we obtained ${\rm C/d.o.f.}=534.6/509$ and the following parameters (1-$\sigma$ errors): $N_H$\footnote{`-l' indicates that the parameter has hit its lower bound.}$=1.9^{+0.6}_{-l}\times10^{20}$~cm$^{-2}$, $\Gamma=1.09\pm0.07$, $L_E=7.3_{-0.5}^{+1.7}$~keV, $L_\sigma=1.5_{-0.3}^{+0.6}$~keV, $L_{st}=7.7_{-2.6}^{+6.6}$, $\alpha=0.56\pm0.25$ and $F=0.91^{+0.03}_{-l}$. Thus, both spectra, from the circle and from the annulus, give similar line parameters and power-law indices (Fig.~\ref{fig:spectra_xspec}). The parameters of the pile-up model also lie in the expected range. The unabsorbed source flux in the 0.3--10.0\,keV range measured using the \texttt{cflux} model corresponds to the luminosity ($1.76\pm 0.18)\times 10^{40}$~erg\,s$^{-1}$ (assuming the distance 6.8~Mpc, \citealt{DrozdovskyKarachentsev2000}).


In order to estimate the significance of the detected line, we refitted the spectrum without the \texttt{gabs} component. This yielded $\rm{C/d.o.f}=595.7/512$ with similar $\Gamma\approx 1.2$ but much different $\alpha\approx0.24$ and $F$ stopped at the limiting value 0.9. Then we repeated the analysis without restrictions for $F$. This improved the fit (583.1/512) but resulted in an unrealistic value of $F\approx 0.84$ with $\alpha\approx 0.13$ and $\Gamma\approx 1.0$. Thus, the addition of a Gaussian absorption to the power-law model improves the fit statistic by at least $\Delta C=48.4$ with only 3 d.o.f difference. This allows us to suspect high significance of the detected absorption feature, however, more stringent estimates will be obtained below, where we will employ a Bayesian approach. 

The shorter observation ObsID\,7086 with an exposure time $\approx 2$\,ks has no source counts above 5.7\,keV. Therefore it was not possible to check for the presence of the absorption line in this observation. We fitted the spectrum extracted from a 2\arcsec\ circular aperture with the \texttt{pileup*tbabs*cflux*powerlaw} model just to measure the source luminosity. The obtained model parameters are $N_H <4.7\times 10^{20}$~cm$^{-2}$, $\Gamma=1.75\pm0.25$ and $\alpha\approx 0.5$ with ${\rm C/d.o.f.}=100.40/110$, and the 0.3--10.0\,keV luminosity is $(7.2\pm 1.2)\times 10^{39}$~erg s$^{-1}$. The addition of the absorption feature with the same parameters as in the longer observation decreases the luminosity by 10\%.

In the \Swift\ observations, as was mentioned above, the source has become much fainter: only 27 counts were accumulated by the standard 30\arcsec\ source aperture during the total exposure time $\simeq 25$~ks, so we combined all the event files to obtain the average spectrum. The source spectrum and counts for the temporal analysis (see below) were extracted from a 30\arcsec\ aperture centred at the ULX coordinates. The background was taken from a large region of about 10 square arcmin area which was a combination of a half-annulus around \obj\ and a rectangular region south in order to not comprise bright sources norths of the ULX. Such a large region was chosen to diminish uncertainty of the background level for a subsequent temporal analysis. Fitting the spectrum with an absorbed power law model, we estimated the ULX unabsorbed luminosity as $(1.7\pm 0.6)\times 10^{38}$~erg s$^{-1}$, which is 50--100 times lower than was in the \Chandra\ observations. The power spectral index and the fit statistic are $\Gamma=1.7_{-0.3}^{+0.8}$ and ${\rm C/d.o.f.}=19.7/19$.

\subsubsection{Bayesian model comparison}
It is known that standard approaches to model comparison like F-test or Likelihood-ratio test are inappropriate for feature-detection problems \citep{Protassov2002}. However, this task can be naturally solved in Bayesian framework. To perform Bayesian calculations, we used the \textsc{bxa} package \citep{Buchner2014bxa} which adapts the numerical library \texttt{UltraNest} for \textsc{xspec}. The models were compared using the Bayes factors. For the observed data $\boldsymbol{y}$ and $i$-th model ${\rm M}_i$ with a parameter vector $\boldsymbol{\theta}_i$, the Bayes factor is $B_{ij}=Z_i/Z_j$, where $Z_i=p(\boldsymbol{y}|{\rm M}_i)=\int p(\boldsymbol{y}|\boldsymbol{\theta}_i)p(\boldsymbol{\theta}_i)d\boldsymbol{\theta}_i$ is called the evidence. The first term in the integral is the likelihood which is $p(\boldsymbol{y}|\boldsymbol{\theta}_i)=\exp{\left[-{1\over 2}C(\boldsymbol{y},\boldsymbol{\theta}_i)\right]}$ in the case of the C-statistic; the second one is the prior distribution of the model parameters. Thus, the Bayes factors are similar to the likelihood ratios but they also take into account parameter uncertainties, while the likelihood ratios compare models only at their best fitting values. Therefore, the conclusion made based on Bayes factors does not depend on a particular set of parameter values. One more advantage of the Bayesian approach is that it includes a penalty for overcomplicated models with an excessive number of parameters. There is a convention to assume a Bayes factor of $B_{12}\ge10$ (or $\log Z_1 - \log Z_2 \ge1.0$) as a sufficiently strong reason to prefer the first model to the second \citep{Jeffreys1961, Buchner2014bxa}.

The Bayesian analysis was done for the ObsID\,9546 spectrum. Besides the two models considered above (a power-law continuum with and without line) we also tested more complicated models: a cut-off power law, a multi-coloured disc blackbody (MCD) and a combination of power-law and multi-coloured disc components, all with and without the absorption line and modified for pile-up and interstellar absorption. Non-informative priors were used for all free parameters; the log-uniform for the normalizations of the continuum components and the uniform distribution on some reasonable interval for the remaining parameters. The intervals were 0.1--12.0\,keV for the line position, 0.0--5.0\,keV for the line width and 0--3 for $\Gamma$. The $N_H$ and $F$ parameters were restricted in the same way as described above. Using \textsc{bxa}, we obtained the posterior distributions of parameters and calculated the log-evidence for each model. 

The results are shown in Table~\ref{tab:spec_results}, where the models are listed in descending order of their log-evidence. Only three models should be considered as acceptable, all of them require an absorption line at $L_{\rm E}\approx 8$~keV. The models with a power-law continuum and with a combination of power-law and disc blackbody continua are almost equally preferred (the disc component contributes about 10\% to the total 0.3--10.0~keV flux), the model with a cut-off power law is a little worse. These three models have shown different spectral indexes $\Gamma$ (still consistent within uncertainties), but the line parameters are about the same. A completely different line energy $L_E\sim 2$~keV is demonstrated by the `disc with line' model, but anyway, this model has to be rejected as well as the models without the absorption line.

\begin{figure}
\includegraphics[angle=270,width=0.49\textwidth]{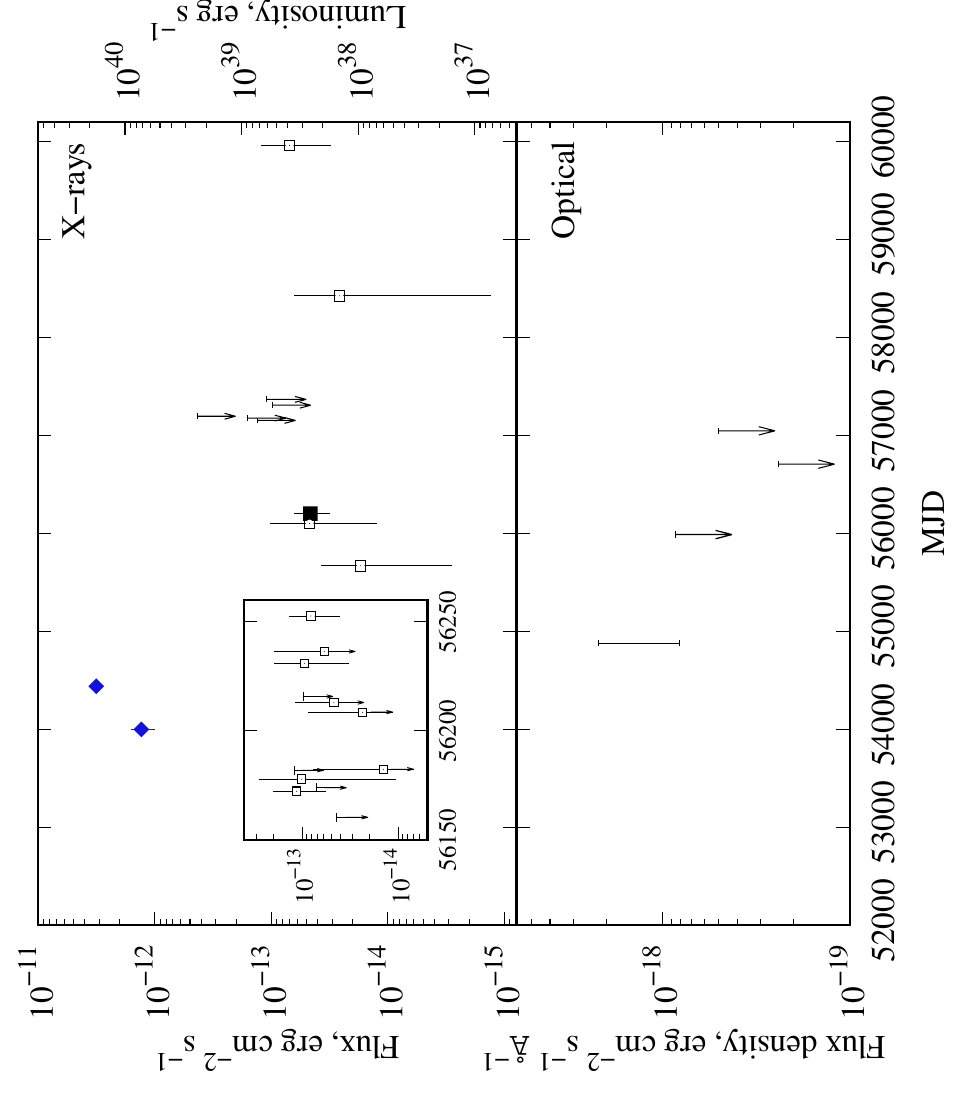}
\caption{Light curve of \obj\ in X-ray 0.3–10~keV band (\textit{top}) and visual B-band (\textit{bottom}). The \Chandra\ and \Swift\ observations are marked by diamonds and squared, respectively. Since the source was very faint in the \Swift\ epoch, we applied the Bayesian technique to obtain the net source count rate in these data (see text). Unfilled squares with error bars denote individual Swift pointings with at least one photon registered, the position of each square marks the mode of the posterior probability distribution of the net flux. Bars mark the shortest 68.3\% credible intervals. If the lower boundary of the credible interval went to zero, we showed it by arrow. For observations with no source photons detected, we estimated 68.3\% upper limits of the net flux and showed them by notches with arrows. Filled black square marks an average of 12 Swift observations carried out in August--November 2012. Inset shows these observations individually. The details of the optical data see in Sec.~\ref{sec:optical}.}
\label{fig:lcurve}
\end{figure}

\begin{figure*}
\includegraphics[angle=0,width=1.0\textwidth]{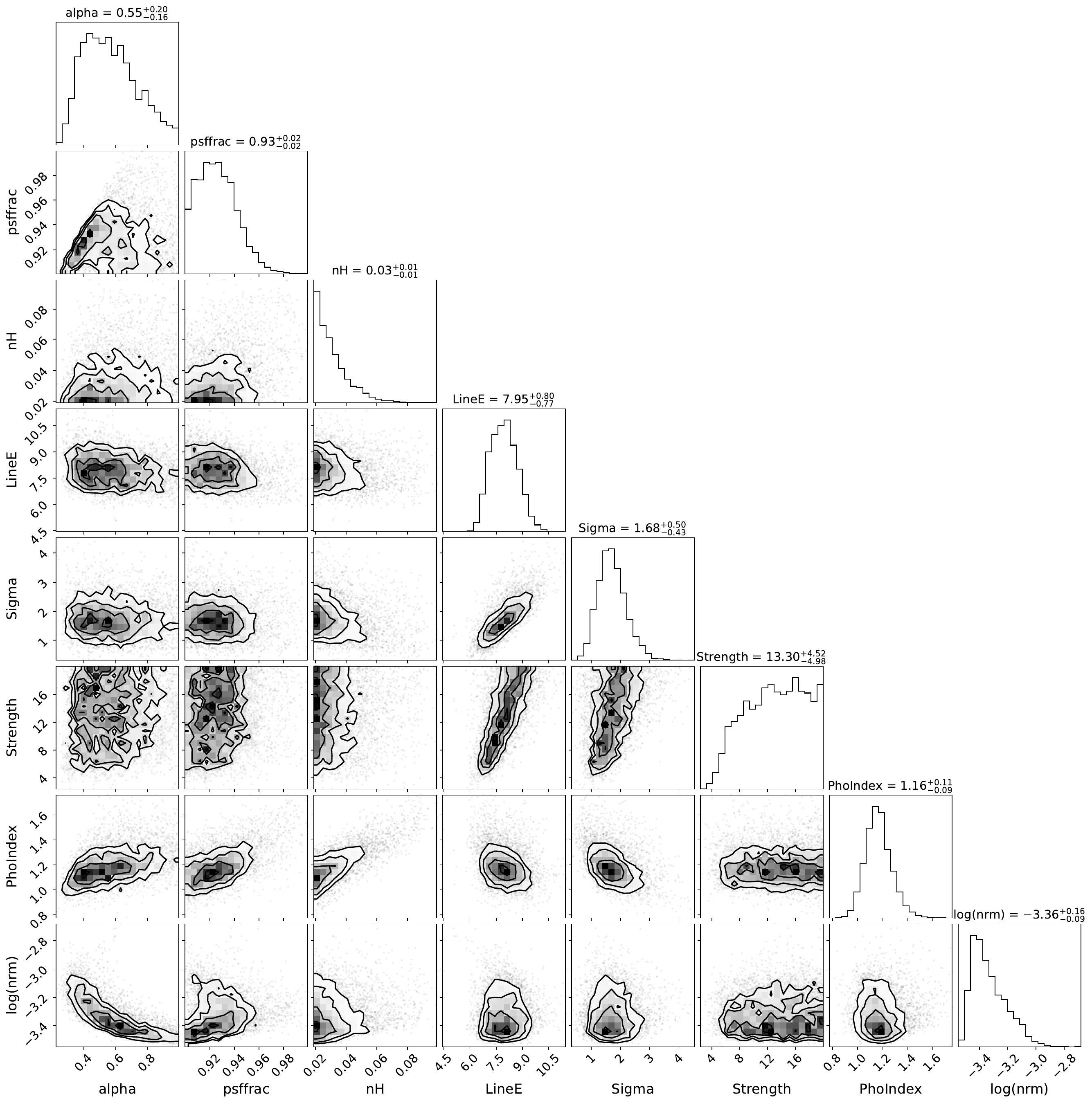}
\caption{One- and two-dimensional posterior probability distributions for the parameters of the \texttt{pileup*tbabs*gabs*powerlaw} model.}
\label{fig:corner_plot}
\end{figure*}

In Fig.~\ref{fig:corner_plot} we present the posterior distributions for the `power law with line' model. It is seen that most of the parameters have a Gaussian-like distribution with a clear maximum which implies that the parameter has an unambiguous best-fitting value. The exception is the line strength whose distribution becomes approximately flat at $L_{\rm st} > 10$ and also shows positive correlations with $L_E$ and $L_\sigma$ (see 2D-distribution in Fig.~\ref{fig:corner_plot}). This is related to the fact that the line is located near the boundary of the \Chandra\ spectral range, so, the more line's blue wing goes outside the range, the deeper and wider the line can be. Besides the posterior distributions, \textsc{bxa} also provides the point in the parameter space where the likelihood maximum is reached. The corresponding line parameters $L_E\approx 7.3$~keV, $L_\sigma\approx 1.5$~keV and $L_{st}\approx 8$ are very close to those obtained by the conventional fitting in the previous subsection.

As for the models without the absorption line, the best result with $\Delta \log Z\approx -1.4$ is demonstrated by the 'PL + Disc` model. However, its very steep power-law component with $\Gamma\sim 3$ tries to fit the soft part of the spectrum. The restriction $\Gamma \leq 2$ in order to make it dominating at high energies significantly decreases the model evidence to $\Delta \log Z\approx -2.5$. It is interesting that the models without the absorption feature included have nevertheless attempted to reproduce it by placing either the Wien tail of the disc blackbody or the exponential cut-off of \texttt{cutoffpl} at the corresponding energy. So, the pure power-law model, which is completely unable to reproduce any local decline, has shown the worst result. This may be considered as an additional argument in favour of that the absorption feature really exists.



\begin{table*}
\begin{minipage}{135mm}
\caption{Swift/XRT observations of \obj.}
\label{tab:swift_data}
\sisetup{range-phrase=--}
\begin{tabular}{cccS[table-format=4]cS[table-format=1.2]S[table-format=2.1]@{\hspace{5mm}}r@{\hspace{1ex}}c@{\hspace{1ex}}l} \hline\hline
{Obs. ID} & {MJD} & {Date} & {${T_{\rm exp}}^a$} & ${N_{\rm src}}^b$ & ${N_{\rm bkg}}^c$ & {\SI[parse-numbers=false]{}{{R_{\rm net}}^d}} &  \multicolumn{3}{c}{\hspace{-5mm}Credible interval$^{e}$}\\ 
	&	 &	& s		 & cnt		& cnt    & {\SI[parse-numbers=false]{10^{-4}}{cnt/s}} &  \multicolumn{3}{c}{\hspace{-5mm}$10^{-4}$\,cnt/s}\\\hline
00031967001	& 55666.6 & 2011-04-15 & 4855 & 3 & 1.30 & 3.4  & 0.6  & $-$ & 7.5 \\
00091496001	& 56098.2 & 2012-06-20 & 1646 & 2 & 0.41 & 9.3  & 2.5  & $-$ & 20.4\\
00091496002	& 56160.3 & 2012-08-21 & 1334 & 0 & 0.14 & 0.0  & 0.0  & $-$ & 8.7\\
00091496003	& 56172.4 & 2012-09-02 & 1264 & 3 & 0.07 & 22.7 & 11.2 & $-$ & 39.7\\
00091496004	& 56173.8 & 2012-09-03 & 814  & 0 & 0.27 & 0.0  & 0.0  & $-$ & 14.1\\
00091496005	& 56177.9 & 2012-09-07 & 402  & 1 & 0.14 & 20.1 & 2.1  & $-$ & 56.4\\
00091496006	& 56181.9 & 2012-09-11 & 477  & 0 & 0.14 & 0.0  & 0.0  & $-$ & 24.1\\
00091496007	& 56182.4 & 2012-09-12 & 1256 & 1 & 0.62 & 2.9  & 0.0  & $-$ & 15.3\\
00091496008	& 56208.5 & 2012-10-08 & 1169 & 1 & 0.41 & 4.7  & 0.0  & $-$ & 17.3\\
00091496009	& 56212.9 & 2012-10-12 & 1151 & 2 & 0.89 & 9.3  & 0.0  & $-$ & 23.6\\
00091496010	& 56215.7 & 2012-10-15 & 597  & 0 & 0.27 & 0.0  & 0.0  & $-$ & 19.3\\
00091496011	& 56230.9 & 2012-10-30 & 919  & 2 & 0.21 & 18.9 & 6.5  & $-$ & 39.0\\
00091496012	& 56236.3 & 2012-11-05 & 527  & 1 & 0.34 & 11.7 & 0.0  & $-$ & 39.1\\
00091496013	& 56252.7 & 2012-11-21 & 2138 & 4 & 0.48 & 16.2 & 8.1  & $-$ & 27.5\\
00084496002	& 57150.8 & 2015-05-08 & 442  & 0 & 0.14 & 0.0  & 0.0  & $-$ & 26.0\\
00084496003	& 57174.1 & 2015-06-01 & 362  & 0 & 0.21 & 0.0  & 0.0  & $-$ & 31.7\\
00084496004	& 57192.7 & 2015-06-19 & 135  & 0 & 0.41 & 0.0  & 0.0  & $-$ & 85.2\\
00084496005	& 57305.8 & 2015-10-10 & 597  & 0 & 0.21 & 0.0  & 0.0  & $-$ & 19.3\\
00084496006	& 57364.9 & 2015-12-08 & 529  & 0 & 0.21 & 0.0  & 0.0  & $-$ & 21.7\\
00084496007	& 58424.2 & 2018-11-02 & 2195 & 2 & 0.82 & 5.2  & 0.3  & $-$ & 12.8\\ 
00084496007 & 59959.9 & 2023-01-15 & 2527 & 5 & 1.44 & 13.9 & 6.2  & $-$ & 24.3\\ \hline
\end{tabular}\\
\textit{Notes.} $^a$GTI corrected exposure time. $^b$Number of total (source+background) counts accumulated in 30'' source aperture. $^c$Number of counts accumulated in the background region and rescaled to the source aperture. $^d$Net source count rate obtained via Bayesian analysis (see text), the mode of the posterior distribution is provided. $^e$ Boundaries of the shortest 68.3\% credible interval of the net count rate.
\end{minipage}
\end{table*}

\subsection{Long-term variability}
\label{sec:xraylc}
Using the flux estimates obtained from the spectral fits (Sec.~\ref{sec:xray_spec}) we have constructed the light curve of \obj. To measure the net (`background subtracted') count rate in the \Swift\ data, we employed the Bayesian approach proposed by \cite{Loredo1992}. Due to the faintness of the source and the shortness of the \Swift\ observations, the individual \Swift\ data sets have accumulated only 2--3 counts in the source aperture or even less, so the direct subtraction of the background level may yield wrong negative values of the net count rate. The Bayesian approach dealing directly with probability distributions helps to avoid this problem. Thus, the posterior distributions of the net count rates were calculated using eq.\,(12.18) from \cite{Loredo1992} which describes the case of Poisson distributed data with (a priori) unknown  background, where we took into account that the source and background measurements have the same exposure time (being obtained from the same images) but different aperture sizes. The results are presented in Fig.~\ref{fig:lcurve} and Table~\ref{tab:swift_data}, where we list our estimates of the net count rate in the form of the modes of its posterior distributions and together with the narrowest 68.3\% credible intervals.  The measured count rates were converted to fluxes using a factor of $5.03\times10^{-11}$~erg~s$^{-1}$~cm$^{-2}$ per 1 cnt/s obtained from the spectroscopy.

It is seen in the light curve (Fig.~\ref{fig:lcurve}) that when the source was observed for the first time in 2006 it had a luminosity of $\sim 7\times10^{39}$~erg\,s$^{-1}$. A year later it was caught at the state about twice brighter. Then, after a break of 3.5 years, \Swift\ found the source to be just above its detection threshold. Although the ULX shows some variability in the \Swift\ data, its luminosity has never exceeded a few times $10^{38}$~erg\,s$^{-1}$ again\footnote{Not taking into account observations of 2015 (near MJD5700 at Fig.\ref{fig:lcurve}) which did not detect any source photons but yielded relatively high upper luminosity limits due to their shortness.}. Moreover, this value should be considered only as an upper limit because the \Swift\ data are contaminated by a nearby source (more likely a background AGN, see Sec.~\ref{sec:extended_obj}), which is located 5.2\arcsec\ away from \obj\ and cannot be resolved by \Swift. We have fitted its spectrum from the longer \Chandra\ observation and obtained $\Gamma = 1.7 \pm 0.5$ with the unabsorbed flux $\approx 1.6\times 10^{-14}$~erg~s$^{-1}$~cm$^{-2}$ (corresponding to $7.5\times 10^{37}$~erg~s$^{-1}$ being brought to the distance of NGC\,5474). This is only twice as low as the mean flux measured by \Swift, so if the AGN is variable, the observed flux in the \Swift\ data, in principle, can be totally provided by it. 
Even so, most of the photons seem to concentrate closer to the ULX position.   

Also, we have searched the longest Chandra observation for pulsations. Using the Python package \textsc{stingray}~1.1.2.4 \citep{Huppenkothen2019OJ,Huppenkothen2019ApJ,stingray112} for the timing analysis, we calculated $Z^2_1$ and $Z^2_2$ statistics for periods from 1000~s to $T/N_{ph}\approx5.6$~s where $T$ is the observation duration and $N_{ph}$ is the number of accumulated photons (the frequency resolution was $\Delta f = 1/T\approx 3.3\times 10^{-5}$~Hz). In both cases we have found no evidences for pulsation. Additionally, we estimated the $3\sigma$ upper limit on the pulsed fraction\footnote{Defined as $2a/(m+a)$ for the signal $r_i=m(1+a\sin(\phi_i))$.} of potentially present periodic signal~--- we obtained 21\% for the highest peak in the $Z^2_1$ spectrum.

\begin{table*}
\begin{minipage}{165mm}
\caption{HST observations, magnitudes and fluxes of the \obj\ optical counterpart.}
\label{tab:HST_data}
\begin{tabular}{ccccccc} \hline\hline
Data Set & Instrument and Filter & $\lambda_{pivot}$ & Total Exp. & Dereddened & Dereddened flux                      & B band mag  \\ 
         &                       &        \AA        &    s     &     mag    & erg\,cm$^{-2}$\,s$^{-1}$\,\AA$^{-1}$ &  interval   \\ \hline
\multicolumn{7}{c}{2009 Feb 9 (MJD~54871)} \\ 
ub9u300[4,5,6,7]m & WFPC2/WF4/F336W & 3344 & 4400 & $22.88\pm0.12$ & $(2.3\pm0.3)\times10^{-18}$ & [24.73, 23.63] \\
ub9u300[1,2,3]m & WFPC2/WF4/F656N   & 6564 & 1800 & $>\,22.1$      & $<\,2.2\times10^{-18}$ & --- \\ 
\multicolumn{7}{c}{2012 Feb 26 (MJD~55983)} \\ 
jbt169010 & ACS/WFC/F606W           & 5922 & 900  & $>\,25.1$      & $<\,2.6\times10^{-19}$ & $>\,24.7$ \\
jbt169020 & ACS/WFC/F814W           & 8046 & 900  & $>\,24.4$      & $<\,1.9\times10^{-19}$ & --- \\ 
\multicolumn{7}{c}{2014 Feb 14 (MJD~56702)} \\ 
icdm43030 & WFC3/UVIS/F275W         & 2709 & 2382 & $>\,25.7$      & $<\,1.9\times10^{-19}$ & --- \\
icdm43040 & WFC3/UVIS/F336W         & 3354 & 1119 & $>\,25.7$      & $<\,1.7\times10^{-19}$ & --- \\
icdm43050 & WFC3/UVIS/F438W         & 4326 & 965  & $>\,26.1$      & $<\,2.5\times10^{-19}$ & $>\,26.0$ \\ 
\multicolumn{7}{c}{2015 Jan 19 (MJD~57041)} \\ 
icnk20020 & WFC3/UVIS/F547M         & 5447 & 554  & $>\,25.7$      & $<\,2.0\times10^{-19}$ & $>\,25.3$ \\
icnk20010 & WFC3/UVIS/F657N         & 6567 & 1545 & $>\,24.7$      & $<\,2.5\times10^{-19}$ & --- \\ 
\multicolumn{7}{c}{2021 Oct 3 (MJD~59490)} \\ 
iec0c3010 & WFC3/UVIS/F218W         & 2228 & 2160 & $>\,23.9$      & $<\,1.3\times10^{-18}$ & --- \\ \hline
\end{tabular}\\
\textit{Notes.} All magnitudes are given in the VEGAmag system.
\end{minipage}
\end{table*}

\begin{figure*}
\includegraphics[angle=0,width=0.6\textwidth]{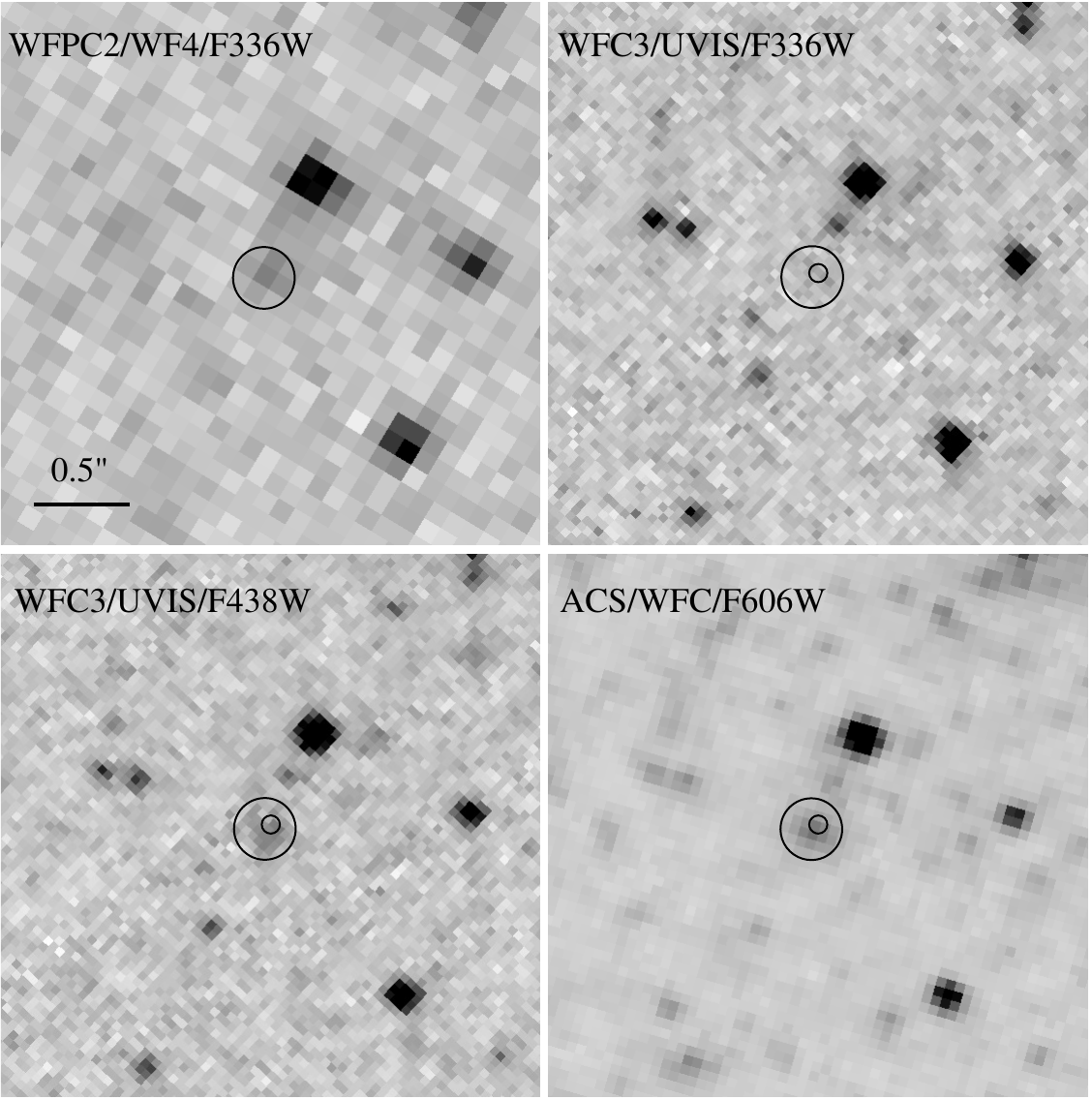}
\caption{Position of the X-1 optical counterpart in different HST images. The big (0.16\arcsec) and the small (0.05\arcsec) circles indicate the corrected \Chandra\ position of the ULX from \protect\cite{Avdan2016} and the position of the point source measured during its outburst in the WFPC2/F336W image, respectively.}
\label{fig:x1pos}
\end{figure*}

\section{Optical data}
\label{sec:optical}
A possible optical counterpart of \obj\ was identified in our previous work \citep{Avdan2016} using images from the \HST\ and \Chandra. In that paper, to obtain shifts between optical and X-ray images, we involved 6 reference sources clearly seen in both bands. As a result, it became possible to reach a very high accuracy of the X-1 position in the \HST\ images (0.16"), which allowed us to unequivocally select a single object. The source was weak (m$_V \simeq $ 24.7), its $(V-I)_0$ colour corresponded to a cold F-G type star. 
However, the source turned out to be slightly extended, about twice larger than surrounding stars, and that became a surprise to us.

While in the paper \cite{Avdan2016} we checked only two images (ACS/WFC/F606W and ACS/WFC/F814W), in the current work we analyse all the available \HST\ observations of the region around \obj. It was observed with the \HST\ once using the Wide Field and Planetary Camera~2 (WFPC2), once with the Wide Field Channel of the Advanced Camera for Surveys (ACS/WFC), and three times with the UVIS channel of the Wide Field Camera~3 (WFC3/UVIS). The log of the HST observations is listed in Table~\ref{tab:HST_data}. Since the object is located in a relatively crowded region (neighbouring more bright stars are only about 0.4--0.5\arcsec\ away from the ULX counterpart), the data from ground-based telescopes cannot be used. 

In the oldest observation taken with WFPC2 in the F336W filter on 2009 February 9 (MJD = 54871), we discovered a bright point source inside the $0.16\arcsec$ error circle. The source is clearly seen in each of four sub-exposures obtained during this observation, but it disappeared in the following \HST\ observations. Comparing images from the WFPC2 and ACS, we have revealed that the point source overlays with the extended object found by \cite{Avdan2016}, but their centres do not coincide (at the confidence level of $2\sigma$); the point source is located on the wing of the extended one. The results of the relative astrometry between the WFPC2 and ACS or WFC3 data is presented in Fig.~\ref{fig:x1pos} where we show four \HST\ images with marked positions of the point source and the $0.16\arcsec$ circle from \cite{Avdan2016}. To apply astrometric corrections we used 3--4 bright isolated reference stars around the object. After accounting for possible shifts, rotation and scaling, the derived standard deviation of the difference between the measured and calculated positions of reference stars became less than 0.3 pixel ($\sim 0.01\arcsec$) for all the WFC3 images and less than 0.2 pixel (also $\sim 0.01\arcsec$) for the ACS images. The resultant uncertainty of the point source position is about $0.05\arcsec$ at 90\% confidence level, where the main part of this quantity is contributed by the error of the source position on the WFPC2 image (the source is located on the WF4 chip having angular resolution of about 0.18\arcsec\ and detected with a signal-to-noise ratio $\approx9$). 
Thus, we can suppose that it is the point source that is the true counterpart of \obj\ caught during its optical outburst, while the extended source is more likely an old stellar cluster (see Sec.~\ref{sec:extended_obj}). 
This idea is also supported by the fact that the \HST\ data set, where the point source is clearly seen in the visual band, is the closest to the \Chandra\ observation in which X-1 was bright in X-rays. Therefore, hereafter saying 'the optical counterpart', we will mean the discovered point source.

To measure the brightness of the \obj\ optical counterpart in the oldest observation, we performed a point-spread function (PSF) photometry. We analysed the {\tt c0f} images using \textsc{hstphot}\,v1.1, which outputs VEGA magnitudes in an infinite aperture corrected for the charge transfer efficiency (CTE). The reddening correction was done in the \textsc{synphot} package using the extinction value A$_V = 0.03$, estimated from the observed Balmer decrement of the nebula surrounding the object (see \citealt{Avdan2016} for more details). 
For the second observation carried out on the same date with the narrow-band filter WFPC2/F656N, we obtained only an upper $3\sigma$ limit to the source magnitude. The results are given in Table~\ref{tab:HST_data}. 

\begin{figure}
\includegraphics[angle=0,width=0.45\textwidth]{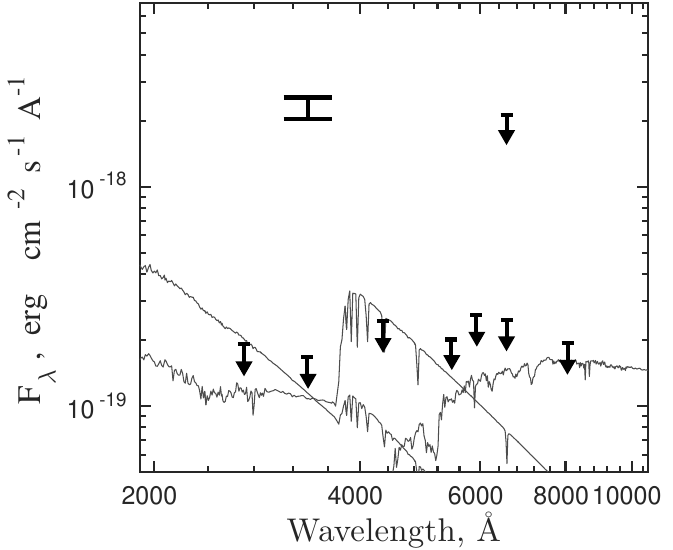}
\caption{Observed flux of the X-1 optical counterpart in the WFPC2/F336W filter and upper limits to the fluxes in the other filters (for the bright and faint states of the source). The model spectra of a B3 main sequence star, an A2 and an M0 bright giants are shown with lines.}
\label{fig:x1sed}
\end{figure}

For the other observations where the counterpart did not stand out over the extended object, we estimated upper limits to its magnitudes performing an aperture photometry with a very small aperture size to decrease contamination from the extended object. The photometry was performed on the drizzled, CTE-corrected ({\it drc}) images using the \texttt{apphot} package in \textsc{iraf}. The apertures were centred at the astrometrically corrected positions of the X-1 counterpart in each image; their size was 1.5~pix. To determine corresponding aperture corrections we used from three to eleven bright isolated stars. The background was measured in a concentric annulus around the objects. The annulus for the WFC3 data was chosen to have an inner radius of 12.5 pixels ($0.5\arcsec$) and an outer radius of 25 pixels ($1.0\arcsec$); for the ACS data, the region had an inner radius of 10 pixels ($0.5\arcsec$) and an outer radius of 20 pixels ($1.0\arcsec$). The results are summarised in Table~\ref{tab:HST_data} and Fig.~\ref{fig:x1sed}. If the measured flux corrected to the infinite aperture appeared to be less than the $3\sigma$ detection threshold in a particular observation, we provided in the table the $3\sigma$ value as an upper limit to the object brightness. We note that the measurements in the most far UV filter F218W gave very high upper limits, significantly higher than the other broadband filters gave. 
We do not use the flux upper limits from this filter in the subsequent analysis because they are unable to provide any additional constraints that would help to clarify the nature of the source.

It is seen (Fig.~\ref{fig:x1sed}) that \obj\ is a highly variable source in the optical range, as well: the F336W filter flux has changed by a factor more than 13 between the observations obtained on two different dates. Since the observations of the remaining dates were conducted with a different filter set (Table~\ref{tab:HST_data}), we converted the WFPC2/F336W, WFC3/F438W, WFC3/F547M and ACS/F606W magnitudes to the Johnson B band using \textsc{synphot}, in order to be able to construct the source light curve. The B band was chosen because it is intermediate for our set of filters, which allowed us to diminish extrapolation errors. To perform such a conversion one needs to know the shape of the spectrum. For observations of the bright state of the counterpart, we approximated fluxes in the F336W and F656N filters by a power law $F_{\lambda} \propto \lambda^{\alpha}$, where $\lambda$ is the pivot wavelength of a filter, and obtained $\alpha \simeq -0.1$. However, since we have only the upper limit to the F656N flux (Table~\ref{tab:HST_data}), the derived value of $\alpha$ should be considered as the upper boundary of the possible spectral slopes; the true spectrum could be steeper. Thus, \cite{Tao2011counterparts} have found that most ULXs have power-law optical spectra with spectral indices of $-3 \div -4$. Therefore, for the lower boundary, we chose $\alpha=-4$ (Rayleigh-Jeans law). Depending on the spectral slope, extrapolation of the WFPC2/F336W flux to the B band yields the range of magnitudes $m_B = [24.73, 23.63]$. 

The F438W filter is close to the B band, so for this filter we obtained that the converted magnitude weakly depends on the chosen slope varying within 0.1~mag. When converting the upper limits from the F547M and F606W filters, we used the Rayleigh-Jeans law to get the most conservative estimates of $m_B$. The magnitudes converted to the B band are also shown in Table~\ref{tab:HST_data} and in the light curve in Fig.~\ref{fig:lcurve}. Also, we have tested the images for possible instrumental variation. To do this, we selected five test stars and obtained their B magnitudes by the same way as we did it for the X-1 counterpart. The spectral slope of each star was determined individually from photometry. In all the cases variations of fluxes of the test stars can be considered as negligible. 

It can be assumed that the optical radiation of the ULX in the low luminosity state comes mainly from the donor star since the star must be a much more persistent source than the accretion disc which produces its optical radiation by reprocessing of X-rays \citep{Vinokurov2013}. Therefore, it should be possible to constrain the donor parameters by 
fitting the observed spectral energy distribution (SED) of the low state with stellar models. We used the ATLAS9 Kurucz models \citep{CastelliKurucz2003Atlas9} with metallicity [M/H]$ = - 0.5$ that approximately corresponds to the metallicity of the galaxy (z = 0.004, \citealt{Sabbi2018}). The luminosity classes were restricted to satisfy the absolute magnitudes $M_V > -3.6$ derived from the observed F547M magnitude with assuming the distance to be 6.8~Mpc. The absolute magnitudes for different classes of stars were taken from \cite{StraizysKuriliene1981StelEvolTracks}. As a result, we found possible spectral classes of the donor star: it can be a main sequence star starting from the B3 class and colder, a star of the III or IV luminosity class from B5 and colder, or a bright giant (II class) from A2 to M0 (Fig.~\ref{fig:x1sed}). This range of possible classes is not narrow because we have only upper limits to the optical flux in the low luminosity state, but,  nevertheless, this analysis has confidently shown that the donor mass must be less than $7M_{\odot}$.

\subsection{The nature of the extended object around the ULX}
\label{sec:extended_obj}

The centre of the extended object is offset from the ULX position by about $0.06\arcsec$. The object is asymmetric, its FWHM size is about $0.23\arcsec \times 0.16\arcsec$ and slightly decreases from the WFC3/F438W to ACS/F606W and F814W images (surrounding stars have sizes of $0.08\arcsec$ and $0.10\arcsec$ in the WFC3 and ACS images, respectively). This corresponds to a linear size of approximately $7 \times 5$ pc at the galaxy distance 6.8~Mpc. In the other filters, the object is much fainter or completely invisible which makes it impossible to investigate its structure.

\begin{figure}
\includegraphics[angle=0,width=0.48\textwidth]{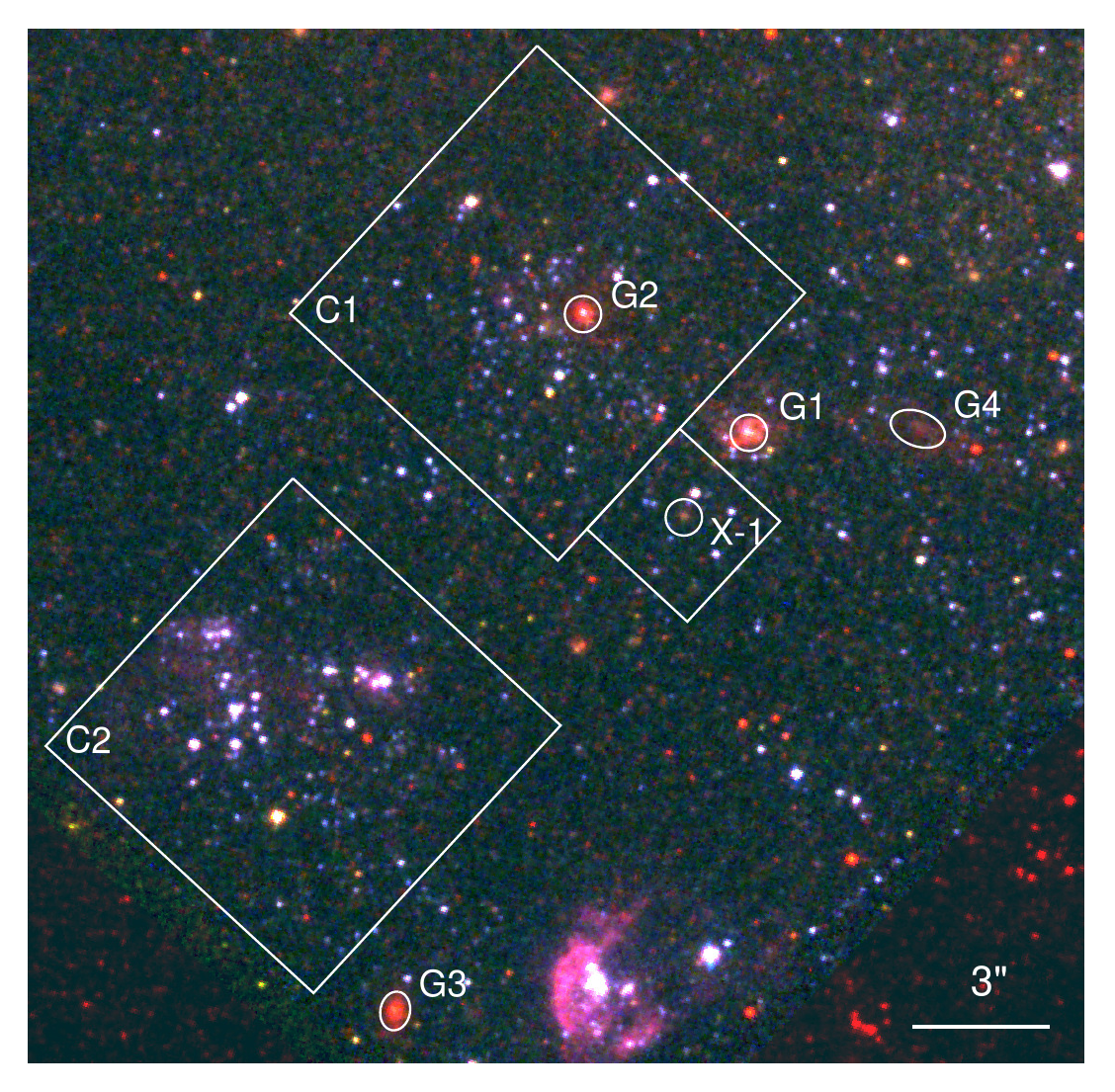}
\caption{HST colour image of the X-1 region (blue: WFC3/F336W; green: WFC3/F438W; red: ACS/F606W). Two squares of 8\arcsec\ and the square of 3\arcsec\ delineate the C1 and C2 groups of young stars and the stars closest to X-1, respectively. Background galaxies at z=0.356 are denoted by G1, G2, G3 and G4 (see text for details). North is up.}
\label{fig:HST_colour}
\end{figure}

Although above we referred to the extended source as a stellar cluster possible hosting the ULX, actually it could also be a nebula or a background galaxy. The presence of four galaxies in the area around the ULX (see the colour HST image, Fig.~\ref{fig:HST_colour}) prompted us to check the nature of the object more carefully. Some of these galaxies (at least G1 and G4 in Fig.~\ref{fig:HST_colour}) look distorted by tidal forces, which suggests that all these galaxies may belong to a single group. The galaxy nearest to X-1 designated as G1 was studied by \cite{Avdan2016}. The authors have identified in its optical spectrum the [O~II] $\lambda$3727 emission line, absorption lines Fe\,{\sc i} $\lambda\lambda$3734,3746, Ca\,{\sc ii}~K and Balmer lines from H11 to H${\delta}$; this allowed them to measure the redshift z~=~0.359. 

The bright nucleus-like source in the centre of the G2 galaxy well coincides with the X-ray source that considered in Sec.~2 as a possible AGN. The remaining residuals between the positions of this X-ray source and the G2 nucleus after applying astrometric corrections does not exceed 0.12$\arcsec$, which is within the uncertainty of the source position in the Chandra image. For the distance z=0.359, the X-ray flux from this source corresponds to a typical AGN luminosity of $6\times10^{42}$~egs/s. The derived X-ray to optical flux ratio is also quite normal for AGN,  we found $\log{(F_X/F_V)} = 0.44$\footnote{We used the relation $\log{(F_X/F_V)} = \log{F_X} + m_V/2.5 + 5.37$, where $F_X$ is the observed X-ray flux in the 0.3\,--\,3.5~keV band and $m_V$ is the visual magnitude \citep{Maccacaro1988galaxiesEinstein}.} while usually observed values are in the range from -1 to 1.7 \citep{Maccacaro1988galaxiesEinstein,Aird2010AGNLxLopt}.

To clarify the nature of the extended object, we performed surface photometry with the {\sc apphot} package on those WFC3 and ACS \texttt{drc} images where the X-1 optical counterpart was not in outburst. We used a circular aperture of $0.175\arcsec$ radius (4.4 and 3.5 pixels for the WFC3 and ACS images, respectively) in order to avoid contamination from the nearest relatively faint star at a distance of $0.3\arcsec$. Since the object is extremely weak in some filters, it was decided to center the aperture by coordinates applying astrometric corrections to each image. The original object coordinates were obtained from the F606W image where the source was detected with the highest signal-to-noise ratio ($S/N \simeq 25$); corrections for other images with respect to the F606W were determined using three bright isolated stars. Since the studied object resides in a quite crowded field, significant variations of the background level can take place. So, to take into account additional uncertainties related to the background subtraction procedure, we have measured the background from three concentric annuli around the object (for the images with relatively good S/N: F438W, F547M, F606W and F814W) with an inner and an outer radii of $0.24\arcsec$ and $0.5\arcsec$, $0.5\arcsec$ and $1.0\arcsec$, $1.0\arcsec$ and $1.5\arcsec$, respectively. The differences between the background levels in these annuli appeared to be minor, they were included into the final errors of the measured magnitudes. For the F336W and F657N images with worse S/N~$\sim 4$, we used only the largest annulus. For the F275W and F218W images where the source is not visible, we estimated a $3\sigma$ threshold for the source magnitude in the $0.175\arcsec$ aperture. The resulting magnitudes corrected for the reddening (A$_V = 0.03$) with \textsc{synphot} are presented in Table~\ref{tab:extobj_photometry}.


\begin{table}
\begin{center}
\caption{VEGA magnitudes and fluxes of the extended source corrected for reddening.}
\label{tab:extobj_photometry}
\begin{tabular}{ccc} \hline\hline
Instrument and Filter & Dereddened & Dereddened flux \\ 
 & mag & erg\,cm$^{-2}$\,s$^{-1}$\,\AA$^{-1}$ \\ \hline
WFC3/UVIS/F218W & $>\,23.7$        & $<\,1.5\times10^{-18}$ \\
WFC3/UVIS/F275W & $>\,25.4$        & $<\,2.5\times10^{-19}$ \\
WFC3/UVIS/F336W & $25.30 \pm 0.27$ & $(2.5\pm0.7)\times10^{-19}$ \\
WFC3/UVIS/F438W & $25.24 \pm 0.10$ & $(5.4\pm0.5)\times10^{-19}$ \\
WFC3/UVIS/F547M & $25.01 \pm 0.15$ & $(3.6\pm0.5)\times10^{-19}$ \\
WFC3/UVIS/F657N & $24.44 \pm 0.26$ & $(3.1\pm0.9)\times10^{-19}$ \\ 
ACS/WFC/F606W   & $24.57 \pm 0.04$ & $(4.25\pm0.18)\times10^{-19}$ \\
ACS/WFC/F814W   & $24.07 \pm 0.05$ & $(2.66\pm0.14)\times10^{-19}$ \\ \hline
\end{tabular}\\
\end{center}
\end{table}

\begin{figure}
\includegraphics[angle=0,width=0.45\textwidth]{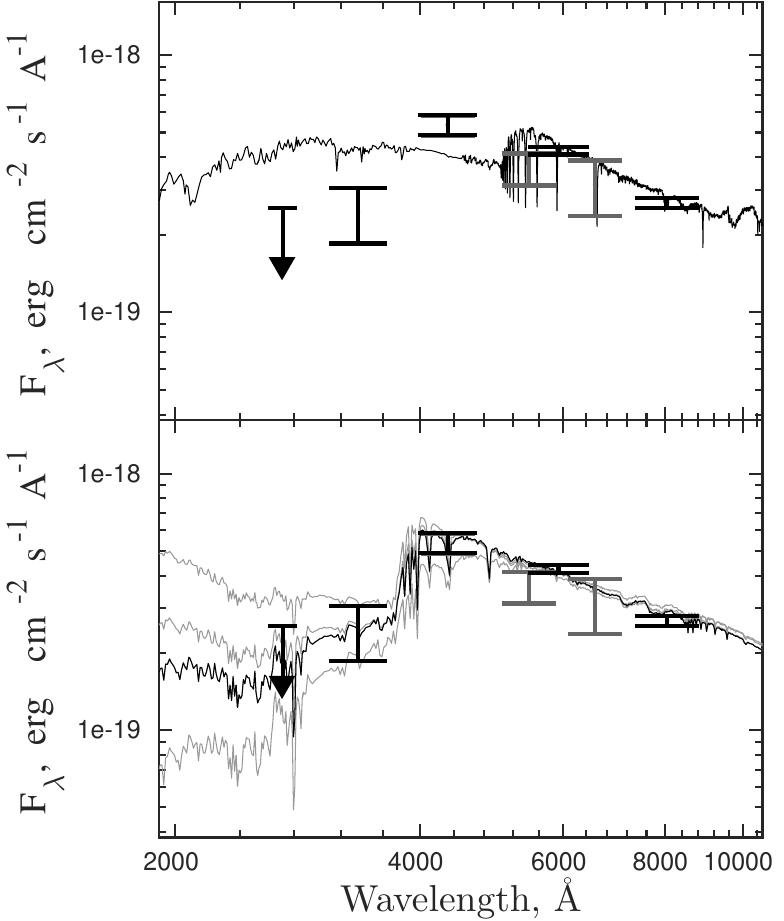}
\caption{Spectral energy distribution of the extended source together with the \textsc{magphys} best fitting model of a galaxy at z=0.359 (top panel) and the \textsc{starburst99} models of star clusters (bottom panel). The best-fitting model spectrum of a 1.0~Gyr age cluster is shown in black, while the spectra for the ages of 0.8, 0.9 and 1.1 Gyr in grey.}
\label{fig:extobj_sed}
\end{figure}

The source shows a SED (Fig.~\ref{fig:extobj_sed}) with a strong rollover below 4000\,\AA. The absence of an excess in the F657N filter whose range covers the H${\alpha}$~$\lambda6563$ and [N\,{\sc ii}] $\lambda\lambda6548,6583$ lines, rules out a nebula as a possible explanation of the source nature. In order to choose between galaxy and star cluster, we have fitted the source SED with the \textsc{magphys} (Multi-wavelength Analysis of Galaxy Physical Properties, \citealt{daCunha2008MAGPHYS}), the \textsc{LePhare} \citep{Arnouts1999LePhare1,Ilbert2006LePhare2} and the \textsc{starburst99} \citep{Leitherer1999Starburst99,Vazquez2005Starburst99OldPop} codes.

The \textsc{magphys} code constructs spectra of galaxies in a wide range of wavelengths using stellar population synthesis models combined with self-consistent models of the dust emission, dust absorption, and the PAH (polycyclic aromatic hydrocarbons) emission. It forms a grid of synthetic models for each observed SED and fit them to the data, which allows to reproduce spectral distributions of many types of galaxies (e.g. \citealt{Brown2014galaxySEDs}). Unfortunately, \textsc{magphys} does not include nebular emission lines, however, we have found that the flux in the F657N filter indicates the absence of emission lines.

We involved only wide-band filters in our analysis. The \textsc{magphys} contains a large library of throughput curves of filters of various telescopes, including all the wide-band filters used in this work excepting WFC3/UVIS/F438W. Therefore, we converted the flux in this filter into the available band ACS/WFC/F435W. Since the difference between the throughputs of these two filters is not very large, the recalculated flux has changed very little and practically does not depend on the particular value of spectral slope used to perform the conversion. Also, \textsc{magphys} requires to assign a redshift to fit the data. We assumed that the object may have the same redshift z~=~0.359 as the galaxy studied by \cite{Avdan2016}.

The obtained agreement between the observed SED and the best fitting model produced by \textsc{magphys} for the adopted redshift appeared to be very poor (Fig.~\ref{fig:extobj_sed}). The model follows the measured fluxes only in two red filters, while the fluxes in the remaining three filters and the rollover below 4000\,\AA\ cannot be reproduced. The achieved fit statistic is $\chi^2 = 27$ for five data points and five main model parameters. This gives a formal number of degrees of freedom equalled to zero which does not allow to use the reduced $\chi^2$ as a goodness-of-fit test. However, it has been shown \citep*{Andrae2010chisq} that the number of degrees of freedom is not strictly defined for complicated nonlinear models: it always lies in the range from 0 to N-1 (where N is the number of data points) and may change during the fitting procedure. So, even in the case of 4 degrees, the obtained $\chi^2$ value corresponds to the p-value of $\approx2\times10^{-6}$ which completely rejects the model. As one of the possible alternatives to the reduced $\chi^2$, \cite{Andrae2010chisq} propose to perform a Kolmogorov-Smirnov (KS) test to assess the goodness of fit, since the normalised residuals `(data--model)/error' have to follow the normal distribution with zero mean and unity variance if the model is true. We performed this test and obtained the value of a KS-statistics of 0.55 which corresponds to the p-value of 0.058.

\textsc{LePhare} has been developed for measuring the redshifts of objects of unknown nature by fitting spectra of galaxies, quasars, stars, etc. to the observed SED. Besides the models, the code also contains a large number of observed spectra of galaxies and active galactic nuclei. Using \textsc{LePhare} we found that our SED (five wide-band filters) could be well fitted by the model of either an active galaxy at z~=~2.68 (the code did not provide the error for this model) or a non-active galaxy at z$ = 2.56 \pm 0.08$; the fit statistics are $\chi^2=1.92$ and 2.86, respectively. These models interpreted the rollover between the F336W and F438W filters as the Lyman break. For these models we also calculated the KS-statistics, the obtained p-values are 0.31 and 0.97. Such a relatively bad p-value in combination with a low $\chi^2$ in the case of the AGN model seemingly indicates significant overfitting.


To check whether the extended object can be a stellar cluster, we used \textsc{starburst99}. For chosen evolution tracks, this code produces model spectra of continuum emission of a stellar cluster as functions of the cluster age and metallicity (the total mass of the member stars is considered only as a normalisation constant and fixed at $10^6$\,M$_\odot$, by default). We employed  the Padova tracks with AGB (asymptotic giant branch) for the metallicity Z~=~0.004 \citep{Sabbi2018}. The initial mass function (IMF) was specified by a Kroupa law with the indices 1.3 for stars of 0.1--0.5\,M$_\odot$ and 2.3 for 0.5--100\,M$_\odot$. We computed a grid of models for cluster ages ranging from 1~Myr to 2~Gyr with a step 10~Myr. Then, convolving each model with the filters throughputs and tuning the normalisation, we calculated both the $\chi^2$ and the KS statistics. We found that the best agreement has been reached by the model with an age of $1.0_{-0.25}^{+0.15}$~Gyr ($1\sigma$ errors) and a total stellar mass $\sim 3000$\,M$\odot$. The age is mainly constrained by the appearing of the Balmer jump in spectra of old clusters which well fits the observed rollover below 4000\,\AA. The $\chi^2$ statistics is 4.3 for five filters and two free model parameters; the p-value of the KS-statistics is 0.98. The best-fitting model spectrum of a 1.0~Gyr age cluster (black curve) and spectra for ages of 0.8, 0.9 and 1.1 Gyr (gray curves) are shown in the lower panel of Fig.~\ref{fig:extobj_sed}. 

\begin{figure*}
\includegraphics[angle=0,width=0.75\textwidth]{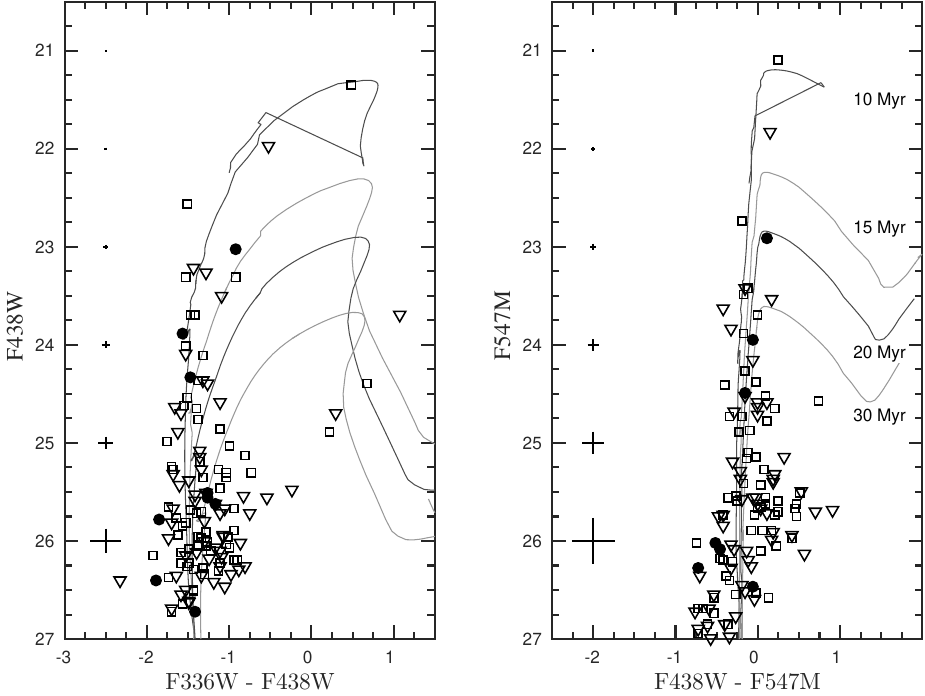}
\caption{Color–magnitude diagrams for the stars of two nearest to \obj\ groups C1 (triangles) and C2 (squares) from Fig.~\ref{fig:HST_colour}. Black circles indicate the positions on CMD of the stars from the small 3" area around X-1. Isochrones of 10, 15, 20, and 30 Myr are shown by lines.}
\label{fig:cmd} 
\end{figure*}

\textsc{starburst99} assumes that the individual properties of the member stars have to be smeared out due to a large number of members and, therefore, the spectral shape does not depend on the total mass. It could be not true for our relatively small cluster. So we decided to conduct an additional test using web-based software \textsc{cmd}\footnote{http://stev.oapd.inaf.it/cgi-bin/cmd}\,v3.3 \citep{Bressan2012StellPopCMD} which simulates a randomised stellar population of a certain age and a certain number of members for the chosen IMF. Besides the parameters of each simulated star, the code can also provide isochrones and photometry in filters of the most popular photometric systems including the HST/ACS and HST/WFC3. Since the population is randomised, the outputs slightly change from one run to another. We used the same canonical two-part power-law IMF as for \textsc{starburst99} and the stellar mass fixed at 3000\,M$_\odot$. Since the outputs include magnitudes separately for each star, we converted all the magnitudes to fluxes and summed them to obtain the total spectrum of the cluster. Eventually, \textsc{cmd} gave a slightly wider spread of acceptable ages than \textsc{starburst99}, from 0.5 to 1.2~Gyr.



So we obtained that the modelling is unable to draw a clear conclusion about the nature of the extended object. Both models: an old cluster of 1.0~Gyr age and a distant galaxy with redshift z~$\approx 2.6$ provide similar quality of the fit. However, the latter interpretation implies that the studied ULX and the background galaxy have coincided by chance. A probability of such a coincidence can be estimated using the Hubble Ultra Deep Field catalogue (UDF catalogue, \citealt{Rafelski2015HubbleUDF}). We have counted about 1000 galaxies in the catalogue with redshifts z~$=2.3\div2.8$ (the $3\sigma$ confidence interval of the redshift measured with \textsc{LePhare}) per the ACS field of view of 11.4 square arcmin. But only two of them have fluxes in the F606W filter equal (within $3\sigma$ errors) to that we obtained for the extended source and 9 galaxies have fluxes equal or greater. The resulting probability that the centre of such a galaxy falls within 0.06$\arcsec$ circle around the ULX is $5\times10^{-7}$ or $2\times10^{-6}$ (for two and nine galaxies respectively), and it will become even less if one takes into account other parameters like the flux in the remaining filters and the colour of the extended source. Thus it is very unlikely that the extended source is a background galaxy; nevertheless, one needs additional observations to make an unequivocal conclusion. If the source is really a galaxy at z~$\approx 2.6$ then one should expect the Balmer jump near 13000\,\AA\ which can be revealed by infrared observation in the filters WFC/IR/F110W and F160W.

In summary, the stellar cluster model fits the observed SED much better than the model of a galaxy with z~$\simeq 0.36$ and comparable with the models of a distant galaxy or an AGN with z~$\simeq2.6$. Accounting for the additional circumstances makes the galaxy interpretation doubtful. The rollover below 4000\AA\ in the observed SED is naturally reproduced by the Balmer jump in the spectrum of an old stellar cluster. The derived total stellar mass and the observed size of about 5~pc do not contradict typical parameters of well-studied low-mass stellar clusters of a few Gyr age residing in our Galaxy \citep{Angelo2019GalClusters}. Thus, we can conclude that the extended object, which practically coincides with the position of the optical counterpart of \obj, is most likely a stellar cluster, whose age is at least several hundred million years.

\subsection{Young stars around \obj}

\obj\ is surrounded by many groups of young stars. Two the nearest big groups denoted as C1 and C2 in the \HST\ colour image (Fig.~\ref{fig:HST_colour}) are located at distances $\simeq200$ and $\simeq300$~pc (about $6\arcsec$ and $10\arcsec$) from X-1, respectively. A small area around X-1 ($3\arcsec$ square in Fig.~\ref{fig:HST_colour}) also contains some relatively bright blue stars, it is possible that they belong to one of the neighbouring groups. 

To determine the ages of the stars in these groups we constructed colour-magnitude diagrams (CMDs). All the stars located within the squares (8\arcsec\ and 3\arcsec, Fig.~\ref{fig:HST_colour}) were assumed to belong to the corresponding group. The magnitudes of the stars were calculated using the PSF photometry, which was carried out on calibrated \texttt{flt}-images with the \textsc{dolphot} 2.0 software \citep{Dolphin2000dolphot} for WFC3. We employed only the data in blue filters (F336W, F438W and F547M), where the majority of old background stars like red giants are not seen. All the necessary tasks ({\texttt{wfc3mask}, \texttt{splitgroups}, \texttt{calcsky} and \texttt{wfc3fitdistort}) were completed, and the parameters recommended in the user’s guide were used to run the \textsc{dolphot}. The magnitudes were derived in the VEGAmag system. Theoretical isochrones for the NGC\,5474 metallicity Z = 0.004 based on \textsc{parsec} v1.2S (Padova and Trieste Stellar Evolution Code, \citealt{Bressan2012StellPopCMD}) were calculated with \textsc{cmd}\,3.3 \citep{Girardi2010GalSurvey} and the extinction A$_V=0.03$ \citep{Avdan2016} was overlaid on them. 

The resultant CMDs, F438W versus F336W-F438W and F547M versus F438W-F547M, are given in Fig.~\ref{fig:cmd}. It is seen that most of the stars concentrate to the upper isochrone, it is seen especially well in the left panel of the figure, F336W versus F438W-F336W. Thus, we can conclude that the youngest stars in these groups are as young as 10 Myr.

\section{Discussion}
Above we have explored the observational properties of \obj. This source demonstrated high variability with a peak X-ray luminosity $\simeq 1.8 \times 10^{40}$~erg/s and a very hard spectrum with $\Gamma\approx 1.05$ in the brightest state. 
Also, we have re-identified the optical counterpart of this ULX and found that it may be located inside a stellar cluster of about a billion years old} (the lower age limit is 0.5 Gyr). Using results of the optical photometry, we applied some constraint on the donor star spectral class and have shown that its mass has to be less than 7\,M$_\odot$. 

The most interesting result is the detection of signs of a roll-off at high energies, which is best described by a Gaussian absorption line with a centroid energy $E_c\approx 8.0$~keV and a width $\sigma\simeq 1.7$~keV. We have shown that despite this line is not seen directly due to pile-up, it is still required to fit the data. Absorption lines of similar relative widths $\sigma/E_c\sim 0.2$ are observed in some X-ray pulsars in Milky Way and Magellanic Clouds and thought to be cyclotron resonant scattering features produced by transitions of electrons between quantized Landau levels in strong magnetic fields in the vicinity of neutron stars \citep{Heindl2004CRSF, Staubert2019CRSF}. The centroid energy of a CRSF is directly proportional to the magnetic field strength $E_c \approx 11.6(1 + z)^{-1}(B/10^{12}\rm{G})$~keV, where the gravitational redshift $z$ is about 0.25 if the emission comes from the neutron star surface. So, the energy $E_c\sim 10$~keV should correspond to the field strength of $\sim 10^{12}$~G in the region of the line formation. 

In the case of ULXs, the cyclotron features have been suspected in three sources: NGC\,300 ULX1 \citep{Walton2018NGC300line}, M51~ULX-8 \citep{Brightman2018M51ULX8NS} and NGC\,4045~ULX \citep{Brightman2022NGC4045CRSF}. However, the latter two sources have the relative line widths much smaller than it is observed in the Galactic pulsars. Thus, for M51~ULX-8, \cite{Brightman2018M51ULX8NS} obtained $\sigma/E_c \sim 0.02$ with $E_c\approx4.5$~keV and interpreted this line as a CRSF produced by proton transitions. Protons being more massive than electrons have lower thermal velocities, but the proton nature of the line also implies three orders of magnitude stronger magnetic fields for the same $E_c$. On the other hand, \cite{Middleton2019M51ULX8} using a more sophisticated spectral analysis method have shown that the wider absorption feature with $\sigma/E_c \simeq 0.2$ still cannot be completely ruled out. In the spectrum of NGC\,4045~ULX, an absorption line of $\sigma\sim 0.1$~keV is seen at $\approx8.6$~keV \citep{Brightman2022NGC4045CRSF}. Because of the high luminosity of the source $L_X>10^{41}$~erg/s, the authors preferred the idea that the absorption feature has an atomic nature (produced by high-level ions of iron blueshifted by $\sim0.25c$), but a proton CRSF is also considered as possible. It is important to note that the detected absorption lines in all the three ULXs does not show any harmonics, whose presence would clearly prove the CRSF nature of those lines.

The relative line width obtained by us more closely matches that of NGC\,300~ULX1. \cite{Walton2018NGC300line}, fitting the cut-off power-law model often used to describe the spectra of pulsar accretion columns to the combined \Chandra+{\it NuStar} spectrum (0.3--30~keV) of the pulsed emission of this ULXP, obtained residuals indicating the presence of a highly significant absorption line with $E_c\simeq 13$~keV and $\sigma\simeq 3$~keV. \cite{Koliopanos2019NGC300noline} have confirmed this result for the same cut-off power-law model, but they also tried another physically motivated model of continuum, `a power law plus two multicolour disc components', for which the addition of the Gaussian absorption line at high energies did not give an improvement of the fit. The latter model, may be more relevant to ULXPs, is based on the assumption that the accretion flow moving along magnetic field lines being captured by magnetic field at the magnetosphere radius $R_m$ has to form an optically thick envelope (so-called `accretion curtain', \citealt{Mushtukov2017NSenvelope,Mushtukov2019timing}) at accretion rates typical of ULXs; the accretion column seen in regular pulsars is thought to be hidden from an observer by this envelope in this case. 
Thus, the first more cold \texttt{diskbb} component in the composite model tested by \cite{Koliopanos2019NGC300noline} represented the accretion disc truncated at $R_m$, the second \texttt{diskbb} was used as a surrogate for the multicolour accretion curtain inside $R_m$, for which there is no specific spectral model in \textsc{xspec}, and the power law (dominating above 15\,keV) described the spectrum of the photons Comptonized in the envelope matter. \cite{Koliopanos2019NGC300noline} have shown that the accretion curtain model is statistically more preferable to the accretion column model, and it does not require the absorption line to describe the NGC\,300~ULX1 spectrum. In our paper, we did not try such a complicated model\footnote{Our `\texttt{diskbb+powerlaw}' model yielded a relatively high temperature $T_{\rm in} \approx 1.6$~keV (Table~\ref{tab:spec_results}) which hints that it is similar to the model by \cite{Koliopanos2019NGC300noline} but without the truncated disc component. Probably, the cold \texttt{diskbb} component with $T_{\rm in} \sim 0.3$~keV is indeed not needed by \obj, because its spectrum looks more straight (Fig.~\ref{fig:spectra_xspec}) and does not have the curvature near 2\,keV that NGC\,300~ULX1 and some other ULXs have. We note that in the case of \obj\ the addition of the Gaussian absorption line has improved the fit for all the tested models, including `\texttt{diskbb+powerlaw}'.} since we have only the \Chandra\ data, which is limited to 10~keV and affected by pile-up, therefore we cannot be completely sure whether the line is really required for our data.

The presence of a CRSF in the \obj\ spectrum implies that the accretor in this ULX must be a highly magnetised neutron star. We have not detected pulsation in the Chandra data whose temporal resolution is limited to 3.2\,s, so this source can potentially be a ULXP with a shorter pulsation period. Actually, the longer periods of tens of seconds (as it is observed in NGC\,300~ULX1,\citealt{Carpano2018NGC300}) also cannot be excluded yet because even the confirmed ULXPs show pulsations only episodically \citep{Bachetti2014Nat,Bachetti2020}. Alternatively, the object might be a `non-pulsing pulsar' where pulsations are invisible for the observer due to, for example, mutual orientation of the magnetic axis, the rotational axis and the line of sight \citep{KingLasota2020}. This scenario is possibly realised in M51~ULX-8 \citep{Brightman2018M51ULX8NS,Middleton2019M51ULX8} which has a strong magnetic field but does not display pulsations.

The indirect signs like the observed high variability amplitude and hard X-ray spectrum also argue in favour of the NS nature of \obj. It is known that the most hard spectra among ULXs belong to ultraluminous pulsars \citep{Pintore2017}. Thus, M82~X-2 shows the spectral index $\Gamma \lesssim 1.3$ at $L_X> 10^{40}$~erg/s, however, the spectrum softens as the source becomes less luminous \citep{Brightman2016M82X2}. The other ULXPs: NGC\,7793~P13, NGC\,5907~ULX, NGC\,300~ULX1, etc. have indices $\Gamma \simeq 1.1\div 1.5$ in the brightest states \citep{Motch2014NaturP13,Israel2017pulsP13,Israel2017SciNGC5907,Carpano2018NGC300}. At the same time, ULXs with no pulsations ever detected show $\Gamma \gtrsim 1.4$ (the hardest of them are Holmberg\,IX~X-1 and IC\,342~X-1, \citealt{Sutton2013NGC5907,Pintore2017}), but at least part of them can actually also be neutron stars. 

Variability amplitude of two orders of magnitude and more, similar to that we have found in \obj, is observed in ULXPs: M\,82~X-2 \citep{Brightman2016M82X2}, NGC\,5907~ULX \citep{Israel2017SciNGC5907}, NGC\,300~ULX1 \citep{Binder2016SN2010daNGC300}. It has been proposed \citep{Tsygankov2016propellerM82} that such a type of variability of ultraluminous pulsars may be related to transitions of the source to the propeller regime in which the magnetic field starts acting as a barrier and blocks the accretion flow if the accretion rate becomes lower than some threshold value \citep{IllarionovSunyaev1975propeller}. However, this mechanism requires magnetar-like magnetic fields ($\sim 10^{15}$~G, \citealt{Tsygankov2016propellerM82, Bachetti2020}) which is, apparently, not the case for \obj. Other possible mechanisms capable of producing variability of high amplitude are: dramatic changes in the accretion rate, for example, due to the motion of the donor star orbit in an elliptical orbit; changes in visibility conditions due to precession of the accretion disc, especially if the emission is highly collimated; instabilities in the accretion disc \citep{Hameury2020,HameuryLasota2020TransientVariability}, etc.

If \obj\ is indeed a highly magnetised neutron star, its residence in an old stellar cluster of about one billion years old may seem unusual because magnetic fields with $B\sim10^{12}$~G are typical of regular radio pulsars which are supposed to be relatively young objects (see, for example, reviews \citealt{KaspiKramer2016PulsarsReview, Igoshev2021NSreview}). Moreover, the vast majority of really old neutron stars with ages of billion years belong to the class of milliseconds pulsars which have magnetic fields of $B\sim 10^{8}\div10^{10}$~G. Actually, such a low field strengths that milliseconds pulsars have are thought to be related to the accretion regime of close binaries at which a low-rate accretion lasts most of their lives and quenches the magnetic field \citep{KaspiKramer2016PulsarsReview}. If a neutron star is isolated or resides in a wide binary system, its magnetic field is subjected only to the `natural' decay. The rate of this decay strongly depends on the internal structure of a neutron star and is still under debate. Some models (for example, \citealt{Kiel2008FieldDecay}) predict that the field weakens by no more than one order of magnitude in the first billion years. Thus, \obj\ might be a binary system, which was non-interacting from the moment of the NS birth until the companion star left the main sequence, and where the accretion began very recently, when the companion filled its Roche lobe. The population synthesis calculation conducted by \cite{Wiktorowicz2017} yielded this scenario as one of the possible evolutionary channels of the ULX formation. It implies that the initial mass of the companion star is $1.3M_\odot$, and the system enters the ULX phase when the donor become either a star of the Hertzsprung gap (400--800~Myr) or a red giant (1200--4300~Myr). The age of the stellar cluster and the upper limit to the donor star mass obtained by us are in a good agreement with this scenario.

Another possibility: the neutron star can actually be young even being a member of an old cluster. Currently, there is known a handful of radio pulsars which show all the signs of their youth (fields $\sim 10^{12}$~G and periods typical of a regular radio pulsars) but, at the same time, they are located in globular clusters \citep{Lyne1996PulsarsinGC,Ransom2008PulsarsinGC}. The most plausible explanation of this phenomenon states that these neutron stars are indeed formed recently, via accretion-induced collapse of a massive white dwarf \citep{Lyne1996PulsarsinGC, Ruiter2019}. 

Finally, \obj\ may be a young object just occasionally projected on the old stellar cluster. Above we have shown that the area around the cluster (including the closest 3\arcsec x 3\arcsec\ region, Fig.~\ref{fig:HST_colour}) is full of stars with ages starting from $\approx10$~Myr (Fig.~\ref{fig:cmd}), so the ULX might be one of them. If the donor star has already left the main sequence, the upper limit to its mass obtained from the photometry $\simeq 7~M_\odot$ suggests that the system is not younger than $\approx40$\,Myr. If the donor mass is less than 3\,M$_\odot$ then the system age is consistent with the age of the old cluster. 

\subsection{Transient ULXs associated with old population}
Not only ultraluminous pulsars exhibit variability amplitudes of several orders of magnitude among ULXs. At present, about two dozens of ULX transients are known \citep{Sivakoff2008,Middleton2012,Pintore2018,Liu2019NGC7090,Earnshaw2019NGC6946,Earnshaw2020NGC925,Brightman2020M51, Dage2021,Walton2021NGC7090}, most of which are less luminous sources that only briefly reach luminosities of $\gtrsim 2 \times 10^{39}$ erg/s during outbursts. In connection with the possible resident of \obj\ in an old stellar cluster, of particular interest are three ULX transients that were reported to likely be associated with old population: M83~ULX-1 \citep{Soria2012M83tULX,Soria2015M83tULX}, M86 tULX-1 \citep{vanHaaften2019M86tULX} and 2SXPS J235825.7-323609 \citep{Brightman2023transients}.

M83~ULX-1 was discovered in the Chandra images taken in December 2010, in which the source was caught at a luminosity close to its maximum (reached in March 2011) of $5\times10^{39}$~erg/s \citep{Soria2012M83tULX}. Since the ULX is not seen neither in previous Chandra images of this area nor in earlier observations by other missions dating back to 1979, it can be concluded that the variability amplitude of this source is of a factor $\sim3000$ or more \citep{Soria2012M83tULX}. In 2014, the source was found to be two times fainter \citep{Soria2015M83tULX}, and later, according to the light curve\footnote{\url{https://www.swift.ac.uk/LSXPS/LSXPS J133705.1-295207}} in the Living Swift-XRT Point Source catalogue \citep{Evans2023LSXPScat}, it slowly decreased its brightness by another three times; the last firm detection occurred in March 2022. X-ray spectra of M83~ULX-1 are more curved than \obj\ has, with a maximum in the range 2--4~keV \citep{Soria2015M83tULX}.

As in the case of \obj, the X-ray outburst of M83~ULX-1 was accompanied by a brightening in the optical range, which allowed identify its optical counterpart \citep{Soria2012M83tULX}. It arose as a blue point source with $M_V = -4.9$ standing out against red emission of surrounding stars. The fact that the HST images taken before the outburst do not contain this blue source suggests that it should be a result of reprocessing of X-rays in our parts of the accretion disc \citep{Soria2012M83tULX}. Analysis of these early HST images allowed to constrain possible types of the donor star~--- the obtained absolute magnitude $M_V \gtrsim -2.1$ and photometric colours correspond to an old ($\gtrsim 500$~Myr) low-mass ($M\lesssim 4M_\odot$) red donor, more likely a red giant or an AGB star~\citep{Soria2012M83tULX}. This behaviour is quite similar to that we found for \obj, where the optical counterpart also was blue and had $M_B = -5.5\div -4.4$ during the outburst  (the scatter is due to uncertainty of the spectral slope, Sec.~\ref{sec:optical}) and $M_B > -3.1$ in the off-state.

The other two sources, 2SXPS J235825.7-323609 and M86 tULX-1, were bright for a much shorter period of time and have no detected optical counterparts. The first one is located at the same area as the pulsar NGC\,7793~P13 which is frequently observed with Swift. The source was firstly detected at its maximum luminosity $\sim3\times10^{39}$~erg/s \citep{Brightman2023transients} but then the flux started falling, and about 180 days later the source became no longer visible by Swift. The XMM-Newton observation carried out $\simeq 210$ days after the initial detection by Swift gave a luminosity $\approx 8\times 10^{37}$~erg/s. All the HST images of this area were obtained during the off-state, so \cite{Brightman2023transients} have not identified any possible optical counterpart to this ULX down to $m_{\rm F814W} \simeq 26$. The weakness of the donor star, along with the fact that the system is located on the galaxy periphery populated by old red giants \citep{RadburnSmith2012NGC7793gal}, indicates that it should be also old. M86 tULX-1 was seen only once, in the Chandra image taken in July 2013, where it had a luminosity of $5\times10^{39}$~erg/s  \citep{vanHaaften2019M86tULX}. The next image of this area was obtained in May 2016 but the source had already faded away. The optical counterpart was not identified, mainly because the galaxy is relatively distant (16.83~Mpc), so the type of the donor is not constrained. The old age of this object was suspected based on that it is located in a giant elliptical galaxy in an area without young stars \citep{vanHaaften2019M86tULX}. There is some possibility for these two objects that they are background/foreground sources, however, their spectra look quite usual for ULXs \citep{vanHaaften2019M86tULX, Brightman2023transients}. 

It is seen that \obj\ indeed shows a great similarity with transient ULXs in old stellar populations, however, their number is too limited to draw the final conclusion. Further studies of such objects will show how well \obj\ fits into the overall picture.

\section{Summary and conclusion}
We have studied the source \obj\, whose peak X-ray luminosity reached $\simeq 1.8 \times 10^{40}$~erg/s. The ULX was bright in two \Chandra\ observation separated by 15 months but the subsequent \Swift\ observations (started 3.5 years later) found it near the Swift/XRT sensitivity limit, i.\,e. fainter by a factor of 50--100 or even more (the \Swift\ data are contaminated by a neighbouring source). Analysing the only good-accumulated X-ray spectrum, we found the signatures of a broad absorption line. We applied a Bayesian approach and have shown that, despite the suffering of the spectrum from pile-up, any tested spectral model requires a Gaussian absorption line component at $E_c\approx 8$~keV to fit the data.

In the optical range, the X-ray source coincides with a red extended object about twice larger than surrounding stars. Analysis of the first HST observation taken about one year after the last observation by Chandra revealed a bright, blue, point-like source within the extended one. We proposed that the blue point source is the true counterpart of the ULX caught during the outburst that lasted at least 29 months (covering both the Chandra and the first HST observations). Then this point source disappeared. The SED of the extended source can be fitted  by the spectrum of either an old stellar cluster of $\gtrsim 500$~Myr with a prominent Balmer jump or a distant galaxy at $z\approx 2.5$, however, the latter option we have considered implausible because of the low probability of X-1 be projected on a background galaxy with such $z$. 
Studying the off-state optical data we derived limits to the possible stellar classes of the donor star: it may be a B3 or colder star of the main sequence, a B5 or colder star of III and IV luminosity classes, or a bright giant (II class) from A2 to M0. In all the cases the star mass must be $\lesssim 7$\,M$_\odot$ which does not contradict the age of the parent cluster.

It the absorption line real, it might be a cyclotron resonant feature, which implies that this ULX hosts a neutron star with a magnetic field $\sim 10^{12}$~G. The hardness of the X-ray spectrum and high variability amplitude, which are the distinctive features of ultraluminous pulsars, support this possibility. In this regard, the fact of the residence of this ULX in an old stellar cluster looks curious, but also an accidental projection of the ULX onto the old cluster cannot be ruled out.

\section*{Acknowledgements}
The authors would like to thank the anonymous referees for useful comments on this paper. We acknowledge the use of public data from the Swift data archive, as well as of data obtained from the Chandra Data Archive, and software provided by the Chandra X-ray Center (CXC) in the application packages CIAO. This research is also based on observations made with the NASA/ESA Hubble Space Telescope obtained from the Space Telescope Science Institute, which is operated by the Association of Universities for Research in Astronomy, Inc., under NASA contract NAS 5–26555. These observations are associated with program IDs 11966, 12546, 13364, 13773 and 16316. The work was performed as part of the government contract of the Special Astrophysical Observatory of the Russian Academy of Sciences (SAO RAS) approved by the Ministry of Science and Higher Education of the Russian Federation.

\section*{DATA AVAILABILITY} 
The data underlying this article will be shared on reasonable request to the corresponding author.

\end{document}